\newcommand{\manuscript}{0}
\newcommand{\userealauthors}{1}

\newcommand{\npatch}{13~}
\newcommand{\npatchtext}{thirteen }
\newcommand{\psSurveyArea}{228}

\newcommand{\ndivtext}{four }

\newcommand{\ncrosstext}{six }

\newcommand{\nautotext}{four }
\newcommand{\nmaskedSources}{108 }
\newcommand{\maxASZ}{1.63}
\newcommand{\bestASZ}{0.63}
\newcommand{\maxSigmaEight}{0.86}
\newcommand{\meanAP}{11.2}
\newcommand{\bestAP}{11.9}
\newcommand{\bestCP}{0.78}

\newcommand{\bl}{{\bm \ell}}
\newcommand{\ave}[1]{\left\langle#1\right\rangle}

\newcommand{\cpuK}{\ensuremath{\times 10^{-5}\,\micro {\rm K}^2}}
\newcommand{\jysr}{\ensuremath{{\rm Jy}^2 {\rm sr}^{-1}}} 
\newcommand{\LCDM}{$\Lambda$CDM}

\newcommand{\be}{\begin{equation}}
\newcommand{\ee}{\end{equation}}
\newcommand{\ba}{\begin{eqnarray}}
\newcommand{\ea}{\end{eqnarray}}

\ifnum\manuscript=2
\documentclass[11pt, letter]{article}

\newcommand{\shorttitle}[1]{}
\newcommand{\shortauthors}[1]{}
\newcommand{\received}[1]{}
\newcommand{\altaffiltext}[2]{#1 #2}
\newcommand{\altaffilmark}[1]{\ensuremath{^{\mathrm{#1}}}}
\newcommand{\keywords}[1]{\emph{keywords}---#1}
\newcommand{\acknowledgements}[1]{ }
\renewcommand{\author}[1]{#1}
\renewcommand{\title}[1]{{\centering {\Large \bf #1}\\ }}

\usepackage[margin=1.2in]{geometry}
\else\ifnum\manuscript=1
\documentclass[12pt,letterpaper,preprint,flushrt]{aastex}
\bibpunct{(}{)}{,}{a}{}{;}
\else
\documentclass{emulateapj}
\fi\fi

\usepackage{graphicx}        
\usepackage{bm}              
\usepackage{natbib}
\usepackage{mathrsfs}

\newcommand{\arone}{148\,GHz}
\newcommand{\artwo}{218\,GHz}
\newcommand{\arthree}{277\,GHz}

\newcommand{\aroneApprox}{150\,GHz}

\newcommand{\commentx}[1]{}

\renewcommand{\vec}[1]{\mbox{\boldmath$#1$}} 

\newcommand{\mat}[1]{\ensuremath{\mathbf #1}}   
\newcommand{\dif}    {\ensuremath{\,\mathrm{d}}} 

\newcommand{\etal}{et al.\,}  

\newcommand{\ra}[3]   
   {\makebox[1.5em][r]{#1}\makebox[1.5em][r]{#2} \makebox[2em][r]{#3}}

\newcommand{\hms}[3]  
   {${#1}^{\mathrm{h}}{#2}^{\mathrm{m}}{#3}^{\mathrm{s}}$}

\newcommand{\hmin}[2]  
   {\ensuremath{{#1}^{\mathrm{h}}{#2}^{\mathrm{m}}}}
   
\newcommand{\hours}[1]  
   {\ensuremath{{#1}^{\mathrm{h}}}}

\newcommand{\dms}[3]  
   {\ensuremath{{#1}\degree{#2}\arcminute{#3}\arcsecond}}

\newcommand{\dm}[2]  
   {\ensuremath{{#1}\degree{#2}\arcminute}}

\newcommand{\ukcmb}  
           {\ensuremath{\micro \kelvin_\mathrm{cmb}}}

\newcommand{\uk}  
           {\ensuremath{\micro \kelvin}}

\newcommand{\fdeg} 
           {\hbox{$.\!\!^{\circ}$}}

\usepackage{color}

\usepackage[mediumspace,Gray,squaren]{SIunits}
\usepackage{rotating}

\shorttitle{ACT Power Spectrum}
\shortauthors{J. Fowler \etal}

\begin{document}

\title{The Atacama Cosmology Telescope: A Measurement of
  the $600<\ell<8000$ Cosmic Microwave Background Power Spectrum at 148\,GHz}

\ifnum\userealauthors=1

\author{
J.~W.~Fowler\altaffilmark{1},
V.~Acquaviva\altaffilmark{2,3},
P.~A.~R.~Ade\altaffilmark{4},
P.~Aguirre\altaffilmark{5},
M.~Amiri\altaffilmark{6},
J.~W.~Appel\altaffilmark{1},
L.~F.~Barrientos\altaffilmark{5},
E.~S.~Battistelli\altaffilmark{7,6},
J.~R.~Bond\altaffilmark{8},
B.~Brown\altaffilmark{9},
B.~Burger\altaffilmark{6},
J.~Chervenak\altaffilmark{10},
S.~Das\altaffilmark{11,1,2},
M.~J.~Devlin\altaffilmark{12},
S.~R.~Dicker\altaffilmark{12},
W.~B.~Doriese\altaffilmark{13},
J.~Dunkley\altaffilmark{14,1,2},
R.~D\"{u}nner\altaffilmark{5},
T.~Essinger-Hileman\altaffilmark{1},
R.~P.~Fisher\altaffilmark{1},
A.~Hajian\altaffilmark{2,1},
M.~Halpern\altaffilmark{6},
M.~Hasselfield\altaffilmark{6},
C.~Hern\'andez-Monteagudo\altaffilmark{15},
G.~C.~Hilton\altaffilmark{13},
M.~Hilton\altaffilmark{16,17},
A.~D.~Hincks\altaffilmark{1},
R.~Hlozek\altaffilmark{14},
K.~M.~Huffenberger\altaffilmark{18},
D.~H.~Hughes\altaffilmark{19},
J.~P.~Hughes\altaffilmark{3},
L.~Infante\altaffilmark{5},
K.~D.~Irwin\altaffilmark{13},
R.~Jimenez\altaffilmark{20},
J.~B.~Juin\altaffilmark{5},
M.~Kaul\altaffilmark{12},
J.~Klein\altaffilmark{12},
A.~Kosowsky\altaffilmark{9},
J.~M.~Lau\altaffilmark{21,22,1},
M.~Limon\altaffilmark{23,12,1},
Y.-T.~Lin\altaffilmark{24,2,5},
R.~H.~Lupton\altaffilmark{2},
T.~A.~Marriage\altaffilmark{2},
D.~Marsden\altaffilmark{12},
K.~Martocci\altaffilmark{25,1},
P.~Mauskopf\altaffilmark{4},
F.~Menanteau\altaffilmark{3},
K.~Moodley\altaffilmark{16,17},
H.~Moseley\altaffilmark{10},
C.~B.~Netterfield\altaffilmark{26},
M.~D.~Niemack\altaffilmark{13,1},
M.~R.~Nolta\altaffilmark{8},
L.~A.~Page\altaffilmark{1},
L.~Parker\altaffilmark{1},
B.~Partridge\altaffilmark{27},
H.~Quintana\altaffilmark{5},
B.~Reid\altaffilmark{20,1},
N.~Sehgal\altaffilmark{21,3},
J.~Sievers\altaffilmark{8},
D.~N.~Spergel\altaffilmark{2},
S.~T.~Staggs\altaffilmark{1},
D.~S.~Swetz\altaffilmark{12,13},
E.~R.~Switzer\altaffilmark{25,1},
R.~Thornton\altaffilmark{12,28},
H.~Trac\altaffilmark{29,2},
C.~Tucker\altaffilmark{4},
L.~Verde\altaffilmark{20},
R.~Warne\altaffilmark{16},
G.~Wilson\altaffilmark{30},
E.~Wollack\altaffilmark{10},
Y.~Zhao\altaffilmark{1}
}
\altaffiltext{1}{Joseph Henry Laboratories of Physics, Jadwin Hall,
Princeton University, Princeton, NJ, USA 08544}
\altaffiltext{2}{Department of Astrophysical Sciences, Peyton Hall, 
Princeton University, Princeton, NJ USA 08544}
\altaffiltext{3}{Department of Physics and Astronomy, Rutgers, 
The State University of New Jersey, Piscataway, NJ USA 08854-8019}
\altaffiltext{4}{School of Physics and Astronomy, Cardiff University, The Parade, 
Cardiff, Wales, UK CF24 3AA}
\altaffiltext{5}{Departamento de Astronom{\'{i}}a y Astrof{\'{i}}sica, 
Facultad de F{\'{i}}sica, Pontific\'{i}a Universidad Cat\'{o}lica,
Casilla 306, Santiago 22, Chile}
\altaffiltext{6}{Department of Physics and Astronomy, University of
British Columbia, Vancouver, BC, Canada V6T 1Z4}
\altaffiltext{7}{Department of Physics, University of Rome ``La Sapienza'', 
Piazzale Aldo Moro 5, I-00185 Rome, Italy}
\altaffiltext{8}{Canadian Institute for Theoretical Astrophysics, University of
Toronto, Toronto, ON, Canada M5S 3H8}
\altaffiltext{9}{Department of Physics and Astronomy, University of Pittsburgh, 
Pittsburgh, PA, USA 15260}
\altaffiltext{10}{Code 553/665, NASA/Goddard Space Flight Center,
Greenbelt, MD, USA 20771}
\altaffiltext{11}{Berkeley Center for Cosmological Physics, LBL and
Department of Physics, University of California, Berkeley, CA, USA 94720}
\altaffiltext{12}{Department of Physics and Astronomy, University of
Pennsylvania, 209 South 33rd Street, Philadelphia, PA, USA 19104}
\altaffiltext{13}{NIST Quantum Devices Group, 325
Broadway Mailcode 817.03, Boulder, CO, USA 80305}
\altaffiltext{14}{Department of Astrophysics, Oxford University, Oxford, 
UK OX1 3RH}
\altaffiltext{15}{Max Planck Institut f\"ur Astrophysik, Postfach 1317, 
D-85741 Garching bei M\"unchen, Germany}
\altaffiltext{16}{Astrophysics and Cosmology Research Unit, School of
Mathematical Sciences, University of KwaZulu-Natal, Durban, 4041,
South Africa}
\altaffiltext{17}{Centre for High Performance Computing, CSIR Campus, 15 Lower
Hope St., Rosebank, Cape Town, South Africa}
\altaffiltext{18}{Department of Physics, University of Miami, Coral Gables, 
FL, USA 33124}
\altaffiltext{19}{Instituto Nacional de Astrof\'isica, \'Optica y 
Electr\'onica (INAOE), Tonantzintla, Puebla, Mexico}
\altaffiltext{20}{ICREA \& Institut de Ciencies del Cosmos (ICC), University of
Barcelona, Barcelona 08028, Spain}
\altaffiltext{21}{Kavli Institute for Particle Astrophysics and Cosmology, Stanford
University, Stanford, CA, USA 94305-4085}
\altaffiltext{22}{Department of Physics, Stanford University, Stanford, CA, 
USA 94305-4085}
\altaffiltext{23}{Columbia Astrophysics Laboratory, 550 W. 120th St. Mail Code 5247,
New York, NY USA 10027}
\altaffiltext{24}{Institute for the Physics and Mathematics of the Universe, 
The University of Tokyo, Kashiwa, Chiba 277-8568, Japan}
\altaffiltext{25}{Kavli Institute for Cosmological Physics, 
5620 South Ellis Ave., Chicago, IL, USA 60637}
\altaffiltext{26}{Department of Physics, University of Toronto, 
60 St. George Street, Toronto, ON, Canada M5S 1A7}
\altaffiltext{27}{Department of Physics and Astronomy, Haverford College,
Haverford, PA, USA 19041}
\altaffiltext{28}{Department of Physics , West Chester University 
of Pennsylvania, West Chester, PA, USA 19383}
\altaffiltext{29}{Harvard-Smithsonian Center for Astrophysics, 
Harvard University, Cambridge, MA, USA 02138}
\altaffiltext{30}{Department of Astronomy, University of Massachusetts, 
Amherst, MA, USA 01003}


\else
 \author{
 Joseph~W.~Fowler,\altaffilmark{1}
 et al.} \noaffiliation
 \altaffiltext{1}{Joseph Henry Laboratories of Physics, Jadwin Hall,
 Princeton, NJ USA 08544}
\fi
\begin{abstract}

  We present a measurement of the angular power spectrum of the cosmic
  microwave background (CMB) radiation observed at \arone.  The
  measurement uses maps with $1.4^\prime$ angular resolution made with
  data from the Atacama Cosmology Telescope (ACT)\@.  The observations
  cover \psSurveyArea\,deg$^2$ of the southern sky, in a 4\fdeg2-wide
  strip centered on declination $53\degree$ South.  The CMB at
  arcminute angular scales is particularly sensitive to the Silk
  damping scale, to the Sunyaev-Zel'dovich (SZ) effect from galaxy
  clusters, and to emission by radio sources and dusty galaxies.
  After masking the \nmaskedSources brightest point sources in our
  maps, we estimate the power spectrum between $600<\ell<8000$ using
  the adaptive multi-taper method to minimize spectral leakage and
  maximize use of the full data set.  Our absolute calibration is
  based on observations of Uranus.  To verify the calibration and test
  the fidelity of our map at large angular scales, we cross-correlate
  the ACT map to the WMAP map and recover the WMAP power spectrum from
  $250 < \ell < 1150$.  The power beyond the Silk damping tail of the
  CMB ($\ell\sim5000$) is consistent with models of the emission from
  point sources.  We quantify the contribution of SZ clusters to the
  power spectrum by fitting to a model normalized to $\sigma_8=0.8$.
  We constrain the model's amplitude $A_\mathrm{SZ}<\maxASZ$ (95\%
  CL).  If interpreted as a measurement of $\sigma_8$, this implies
  $\sigma_8^\mathrm{SZ}<\maxSigmaEight$ (95\% CL) given our SZ model.
  A fit of ACT and WMAP five-year data jointly to a 6-parameter \LCDM\
  model plus point sources and the SZ effect is consistent with these
  results.

\end{abstract}


\section{INTRODUCTION}

\setcounter{footnote}{0}

The cosmic microwave background (CMB) radiation captures a view of the
universe at only $\sim400,000$ years after the Big Bang.  The angular power
spectrum of temperature anisotropies in the CMB has been crucial in
developing the current standard cosmological model, in which the
universe today contains some 5\% baryonic matter, 23\% dark
matter and 72\% dark energy.  We refer to this model throughout as the
\LCDM\ model.  The temperature power spectrum has been measured to
good precision for multipole moments $\ell\lesssim3000$.  At angular
scales with $\ell\lesssim2000$, the power spectrum matches the
predictions of \LCDM, and it can be used to constrain multiple
parameters of the cosmological
model \citep[e.g.,][]{dunkley/etal:2009, brown/etal:2009, 
reichardt/etal:2009, sievers/etal:prep}.  The agreement between the
current polarization anisotropy measurements and the predictions
of \LCDM\ cosmology further supports the model
\citep[e.g.,][]{dunkley/etal:2009,brown/etal:2009,chiang/etal:prep}.

At $\ell\gtrsim 3000$, the signal from the primary CMB anisotropy
becomes dominated mainly by two populations.  The first consists of
point source emission from both radio and dusty infrared-emitting
galaxies.  The second is the population of massive galaxy clusters
that give rise to the Sunyaev-Zel'dovich (SZ)
effect~\citep{sunyaev/zeldovich:1970}, in which CMB photons scatter
off the electrons of the hot intra-cluster medium.  Removal of the
brightest foreground galaxies and SZ clusters can reduce their
contributions to the power spectrum, but the net anisotropy power due
to unidentified sources still dominates the exponentially falling
primary anisotropy spectrum at small scales.

\begin{figure*}[htb]
  \centering
  \resizebox{.9\textwidth}{!}{
  \plotone{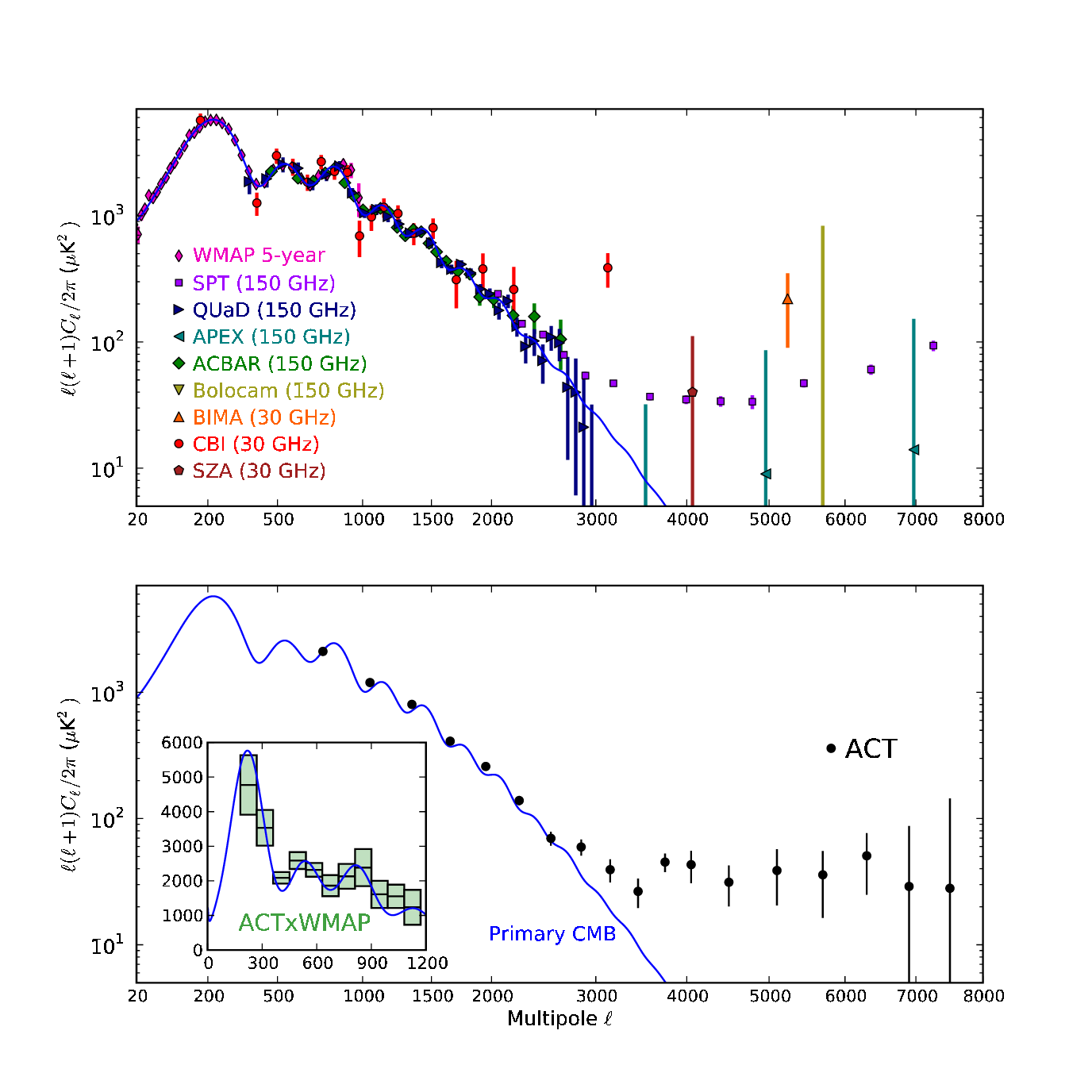} 
} 
\caption{ Recent measurements of the CMB power spectrum, including
  this work.  \emph{Top:} the measurements of WMAP
  \citep{nolta/etal:2009}, Bolocam \citep{sayers/etal:2009}, QUaD
  \citep{brown/etal:2009, friedman/etal:2009}, APEX-SZ
  \citep{reichardt/etal:2009a}, ACBAR \citep{reichardt/etal:2009}, SZA
  \citep{sharp/etal:prep}, BIMA \citep{dawson/etal:2006}, CBI
  \citep{sievers/etal:prep}, and SPT \citep{lueker/etal:prep}.  For
  all the results, a radio point source contribution has been removed
  either by masking before computing the 
  power spectrum (at 150\,GHz), or by masking and modeling the
  residual (at 30\,GHz and for WMAP).  APEX-SZ additionally masks
  clusters and potential IR sources. 
  \emph{Bottom:} The ACT power spectrum from this work. The inset
  shows the cross-power spectrum between ACT and WMAP maps in the ACT
  southern field (see Section \ref{sec:wmap_cal}), which we use to
  check both the validity of the maps at larger scales and the
  absolute calibration.  Only the ACT power spectrum is analyzed in
  this paper.  In both panels and the inset, the solid curve (blue) is
  the \LCDM\ model of \citet{dunkley/etal:2009} (including lensing).
  The SZ effect and foreground sources are expected to contribute
  additional power, as shown in Figure~\ref{fig:cl_binned} and
  Table~\ref{tab:cls}.  For display purposes---and only in this
  figure---we scale our result by 0.96 in temperature relative to the
  Uranus calibration; this calibration factor best fits our data to
  the \LCDM\ model and differs from the Uranus calibration by
  $0.7\sigma$.  Recent WMAP observations of Uranus suggest the same
  rescaling factor (see footnote to Section \ref{sec:calibration}).
  ACT bandpowers for $\ell>4200$ have been combined into bins of
  $\Delta\ell=600$ for this figure; they are given in a note to
  Table~\ref{tab:cls}.  }
  \label{fig:spectrum_many}
\end{figure*}

The top panel of Figure \ref{fig:spectrum_many} shows recent
measurements of the anisotropy at $\ell>2000$. In analyses of 30\,GHz
data, radio point sources are masked out and a residual component is
modeled and subtracted.  For 150\,GHz analyses, radio sources are
masked and the residual radio and dusty galaxy contribution is
estimated.

The SZ effect has a unique frequency signature. In CMB
temperature units (the units of Figure~\ref{fig:spectrum_many}) the
amplitude at 150\,GHz is roughly half that at 30\,GHz. The amount of
SZ power is governed by $\sigma_8$, which measures the amplitude of
the cosmic matter power spectrum on $8h^{-1}$\,Mpc scales; the SZ
power scales approximately as $\sigma_8^7 (\Omega_b h)^2$
\citep{seljak/burwell/pen:2001,komatsu/seljak:2002}. At 30\,GHz, CBI
reports~\citep{sievers/etal:prep} excess emission above the \LCDM\
model at $\ell\approx 3000$ after accounting for all known radio
sources~\citep{mason/etal:2009}.  A possible source of the ``CBI
excess'' is the SZ effect.  The SZA data~\citep{sharp/etal:prep}, also
at 30\,GHz and also after accounting for point sources, are consistent
with the \LCDM\ model at $\ell=4000$. The ACBAR
results~\citep{reichardt/etal:2009} at 150\,GHz are consistent with
the CBI excess and \LCDM.  \citet{sievers/etal:prep} show that all the
above data in the top panel of Figure~\ref{fig:spectrum_many} are
consistent, within 95\% CL, with \LCDM\ plus a SZ contribution of
$\sigma_8=0.922\pm0.047$ ($1\sigma$ error bars).  Recently, the
South Pole Telescope (SPT) group reported $\sigma_8=0.773\pm0.025$
based on the power spectrum at 150 and 220\,GHz
\citep{lueker/etal:prep}.

In this paper, we present a new measurement of the CMB anisotropy
power spectrum in the range $600<\ell<8000$ (corresponding to angular
scales of approximately 1.4\arcminute\ to 18\arcminute).  The
observations were made at \arone\ in 2008 with the Atacama Cosmology
Telescope.  Figure~\ref{fig:spectrum_many} (bottom panel) shows the
results, which are discussed in Section \ref{sec:params}.  We briefly
describe the instrument, the observations, the calibration of the
data, and the map-making procedure.  The dynamic range of the power
spectrum is large enough that we employ new techniques in spectral
estimation.  We confirm that difference maps of the data give power
spectra consistent with no signal and that the power spectrum is
insensitive to several details of our analysis.  We use the shape of
the power spectrum to bound the dusty galaxy contribution and the
power from SZ clusters.

\section{INSTRUMENT AND OBSERVATIONS}
The Atacama Cosmology Telescope (ACT) is a 6-meter, off-axis Gregorian
telescope optimized for arcminute-scale CMB anisotropy measurements
\citep{fowler/etal:2007, hincks/etal:prep}.  It was installed at an
elevation of 5190\,m on Cerro Toco\footnote{ACT is at
  22.9586\degree\ south latitude, 67.7875\degree\ west longitude.}  in
the Atacama Desert of northern Chile in March 2007.  Observing
conditions in the Atacama are excellent owing to the elevation, the
arid climate, and the stability of the atmosphere.  After all cuts,
the median precipitable water vapor (PWV) was 0.49\,mm during the
observations presented here.

The Millimeter Bolometer Array Camera (MBAC), the current focal-plane
instrument for ACT, uses high-purity silicon lenses to reimage
sections of the Gregorian focal plane onto three rectangular arrays of
detectors.  The arrays each contain 1000 transition edge sensor (TES)
bolometers.  Their spectral coverage is determined by metal-mesh
filters \citep{ade/etal:2006} having measured band centers of
\arone, \artwo, and \arthree.  The bolometers are cooled to 300\,mK by
a two-stage helium sorption fridge backed by commercial pulse-tube
cryocoolers.  The telescope performance and control
systems~\citep{hincks/etal:2008, switzer/etal:2008}, camera
design~\citep{thornton/etal:2008, swetz/etal:2008}, detector
properties~\citep{zhao/etal:2008,niemack/etal:2008}, and readout
electronics~\citep{battistelli/etal:2008} are described
elsewhere.\footnote{The site {\tt
    http://www.physics.princeton.edu/act/} archives papers by the ACT
  collaboration.}

\subsection{Observations}
\label{sec:observations}


ACT operated with all three arrays from mid-August to late December,
2008 for the data presented here.  The observing time was divided
between two regions away from the galactic plane.  The deepest
observations cover 900\,deg$^2$ of the southern sky in a strip
$8\degr$ wide centered on $\delta=-53\degr$ with RA from \hours{19} to
\hours{24} and \hours{0} to \hmin{7}{36}.  The analysis presented in
this paper uses only data from the central \psSurveyArea\,deg$^2$ of
the southern strip and only observations made with the \arone\ array.
The slightly elliptical beam has full-widths at half-maximum (FWHM) of
1.40\arcmin\ by 1.34\arcmin\ at this frequency
\citep{hincks/etal:prep}.

The observations were made by scanning the sky at a constant elevation
of $50\degr$.  Each scan is $4\fdeg5$ wide on the sky ($7\fdeg0$ in
azimuth angle).  Scans repeat every 10.2\,s.  Each half-scan consists
of 4.2\,s of motion at a constant speed of 1\fdeg5\,/s followed by
0.9\,s of acceleration.  The first half of each night is spent
observing the field rising in the eastern sky, after which ACT turns
to the western sky to observe the same field as it sets through the
standard elevation of $50\degr$.  The scan strategy is designed to
minimize changes in the telescope's orientation with respect to the
local environment while ensuring cross-linked observations in
celestial coordinates.  Sky rotation ensures that all detectors sample
all points in the field each night, apart from small areas at the
edges.

As the telescope scans in azimuth at constant elevation, each detector
is sampled at 399\,Hz. The data sampling, position reading, and all
housekeeping data are synchronized by a shared 50\,MHz
clock; absolute times are referenced to a GPS receiver with 0.25\,ms
accuracy. The data are stored in continuous fifteen-minute segments
called time-ordered data sets (TODs).  Each TOD requires
1.6\,GB of storage per detector array, or 600\,MB after applying
lossless compression.

In addition to the main survey, we perform occasional calibration
measurements.  Most nights, a few minutes are used to measure a planet
when it passes through our standard elevation.  Any given target is
used only once in every three nights to minimize non-uniformity in the
coverage of the CMB regions.  The planetary observations allow us to
measure the system's relative and absolute responsivities, pointing,
and beam profiles; they also provide a way to check for any time
variations in each.  A series of tuning and biasing procedures are
also followed each night before and after regular CMB observations, to
optimize and roughly calibrate the detector response.

\section{DATA REDUCTION AND MAP-MAKING}

\begin{figure*}[htb]
  \centering
  \resizebox{\textwidth}{!}{
  \plotone{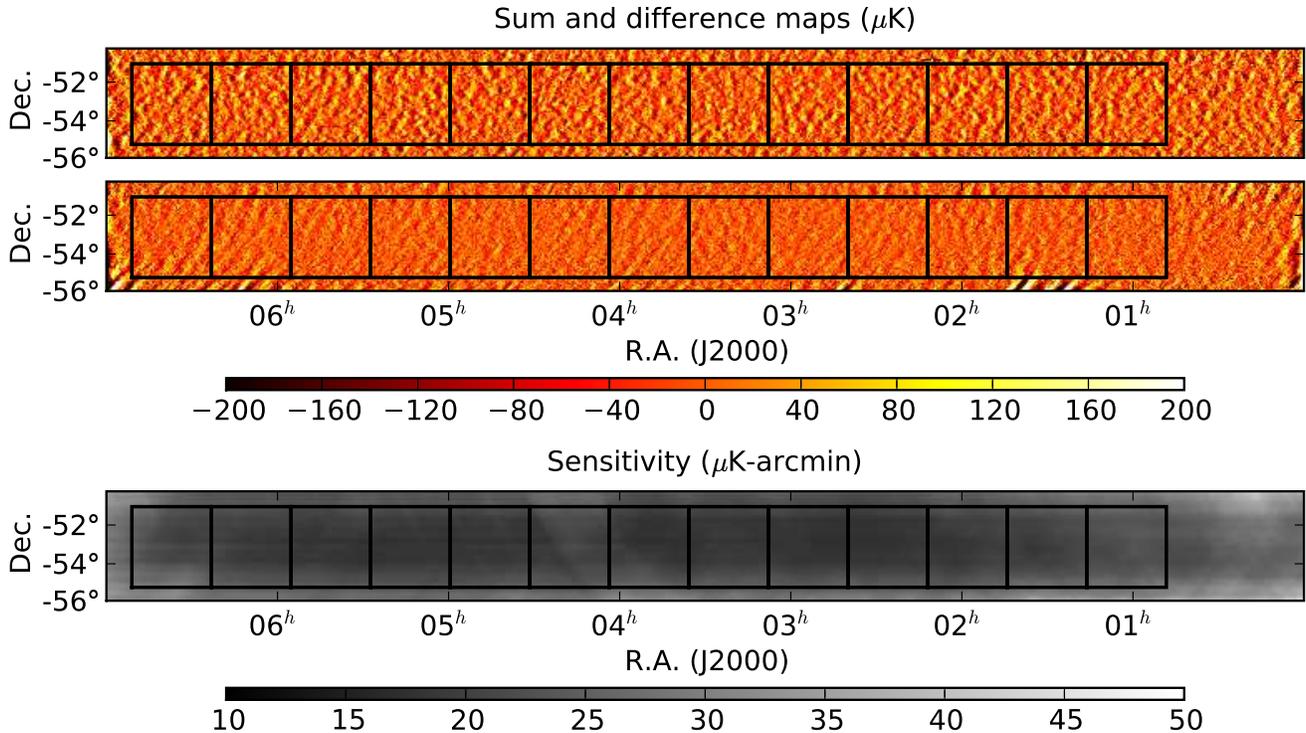} 
  }
  \caption{
    The map and a difference map of the ACT southern field at \aroneApprox.
  The same filters used in the power spectrum analysis are applied to
  both maps: an isotropic high-pass filter suppresses power for
  $\ell\lesssim 300$
  (Equation~\ref{eq:highpass}), and all modes with
  $|\ell_x|<270$ are set to zero, as described in
  Section~\ref{sec:spectrum_single}. 
  \emph{Top:} The ACT southern field.  The intensity scale 
    is \uk\ (CMB units).  The $360\,\mathrm{deg}^2$ with lowest
    noise are shown.  The squares 4\fdeg2 on a side indicate
    the \npatch patches used for the CMB power spectrum
    analysis (\psSurveyArea\,deg$^2$ total).
  \emph{Middle:} A difference map made from two halves of the same
    data set.  Most of the remaining structure visible at large scales
    is well below the range of $\ell$ that we consider in the power
    spectrum analysis. 
  \emph{Bottom:} The rms temperature uncertainty for one square
  arcminute pixels.  
}
  \label{fig:maps}
\end{figure*}

The goal of the data reduction is to estimate the maximum-likelihood
map of the sky.  We select properly tuned detectors and calibrate
their pointing and relative gains.  The absolute gains are determined
by observations of Uranus.  The map-making itself is an iterative and
computationally intensive process.

\subsection{Data selection}

The first step in the data reduction is to select TODs when the
receiver is cold enough for stable operation and when the precipitable
water vapor (PWV) is less than 3\,mm.  Of the $\approx 2880$\,h
between Aug 25 and Dec 24, 2008, MBAC was on line for 1352\,h, roughly
corresponding to the night time hours. After cutting on PWV and 
instrument performance, 1031\,h (35\% of calendar time) went
into the pipeline.  Of those, 850\,h were spent observing the southern
strip.

In the second step of the data reduction, we check whether each
detector's data exhibit problems that warrant removal from the final
analysis \citep{dunner:2009}.  In the case of rare and transient
effects, data are removed for a single detector up to a few seconds at
a time; other problems are diagnosed in 15-minute intervals, and when
necessary, each detector's data may be removed for the entire TOD.  We
check the detector's response speed; look for transient effects such
as cosmic ray hits; and ensure that the feedback loop remained locked,
keeping the SQUID amplifiers in the linear
regime~\citep{battistelli/etal:2008}.  
Finally, we compare the detector's response to atmospheric emission
with that of the array-wide average, in each case filtering out data
at frequencies above 50\,mHz.  Detectors are cut if their data do not
have at least a 0.98 correlation with the average or if the amplitude
of the atmospheric signal is not within 15\% of the median amplitude
found among good detectors.
This last cut is the largest and typically removes 100
detectors. Additionally, we bin the data by scan and compute the rms,
skewness, and kurtosis in each bin. A cut on rms removes an additional
30 detectors on average, while the combined cut on skewness and
kurtosis removes only one additional detector.  When the instrument is
well-tuned, ACT typically has 680 science-grade detectors at
\aroneApprox. The remainder are discarded from the analysis or used to
monitor instrumental effects.  The effective sensitivity of the array,
including all instrumental and atmospheric effects, is
$\sim30\,\uk\,\second^{1/2}$ in CMB temperature units.

\subsection{Calibration to planets}
\label{sec:calibration}

After we have a reliable set of detectors to use, the next step is
gain calibration.  In practice, the cuts are computed twice, so that
the final set of cuts is based on calibrated data.  We use Uranus as
our absolute reference standard.

The atmosphere is an excellent continuous flat-fielding calibrator.
We find that the time variation in detector relative responsivities is
not significant.  Therefore, the detectors' relative responses to the
atmosphere (averaged over the entire season) are used to convert each
output to a common scale.  Observations of Saturn confirm these
relative calibrations;  it is the only source observed in 2008 bright
enough to use for this confirmation.

We take the brightness temperature of Uranus at \arone\ to be
${T_\mathrm{U}= 112\pm 6}\,\kelvin$ \citep{griffin/orton:1993,
  marten/etal:2005, kramer/moreno/greve:2008}.  All results in this
work scale in proportion to this temperature.  For planet
calibrations, we use the beam solid angle $\Omega_A=218\pm4$\,nsr
\citep{hincks/etal:prep}.  The net calibration error of 6\% in
temperature is the combination of these two uncertainties and is
dominated by uncertainty in the brightness of Uranus.


Saturn was used as a rough check on the absolute calibration result.
The Saturn observations suggest that the ACT data are 10\% brighter
than inferred from the Uranus data.  We attribute this inconsistency
to the difficulty in modeling the brightness of the planet and its
rings as they vary over time and across frequencies.

We track the stability of the system over time in a number of ways.
The absolute celestial calibration is checked through measurements of
a planet on most nights. The conversion between raw data and units of
power absorbed on the detectors is calibrated in two ways.  The
conversion is estimated twice nightly by modulating the TES detector
bias voltage with a small additional square-wave and measuring the
response.  A calibration is also performed at the beginning of each
night by sweeping all detectors through the 
full range of bias voltages.  Both methods are described further in
\citet{niemack:2008} and \citet{fisher:2009}.
Based on several tests \citep{switzer:2008}, we find that the gain of
each detector is constant over the season to better than 2\%.

As part of the calibration, the detector temporal response is
deconvolved from the detector time streams.  We model the response
with a single time constant.  The detectors' time constants are
determined from the planet measurements and bias steps
\citep{hincks:2009}.  The median time constant is
$\tau_\mathrm{det}=1.9$\,ms (thus $f_{3\mathrm{dB}}=84$\,Hz), which
corresponds at the ACT scan rate to $\ell\approx 31,000$. Slow
detectors with $f_{3\mathrm{dB}}\le15$\,Hz are not used.  We also
deconvolve the anti-aliasing filter imposed in the data acquisition
system \citep{battistelli/etal:2008}.  We then further filter and
sample the time stream at half the raw rate to speed the map-making
step. This last anti-aliasing filter acts only at angular scales
smaller than the beam and does not affect the maps.

\subsection{Pointing reconstruction}
\label{sec:pointing}

The planet Saturn is bright enough that each bolometer detects
it with high signal-to-noise in a single scan.  We fit in the
time domain to find the best two-dimensional location for each
detector relative to the notional array center.  The fit produces
altitude and azimuth offsets for each, along with information about
the relative gains, the beam sizes, and the detector time
constants.  The detector pointings are consistent with optical models
of the telescope and reimaging optics~\citep{fowler/etal:2007}.  The
relative pointing used in this work is an average over 22 observations
of Saturn.  The rms uncertainty is 1.2\arcsec\ for the detector relative
pointings, which contributes negligibly to any pointing error in the
sky maps.

The location of the array center can also be found from each
observation of Saturn.  It has a scatter of 4.3\arcsec\ rms over
the 22 observations, which we attribute to slight thermal deformations
in the telescope mirrors and their support structures.

The Saturn data give the absolute location of the array center only
for other observations taken at the same horizon coordinates.  To
determine the absolute location of the array center during science
observations, we used approximately 20 known radio sources found in
preliminary maps.  We find this approach both simpler and better
constrained than making a complete pointing model of the telescope.
The ACT southern field was observed both rising and setting, at
azimuths centered at 30\degree\ on either side of south.  These
preliminary maps were made separately for rising and setting data and
were used to determine separate pointing corrections for the two
cases.  After correction, we estimate the maps to have 5\arcsec\
pointing uncertainty.

\subsection{Map-making}
\label{sec:mapmaking}

The goal of the map-making step is to take the 3200~GB  
of cut, calibrated, and deconvolved raw data from both the rising and
setting scans and produce from it a maximum likelihood estimate of the
sky.  This work produces a 200~MB map, 16,000 times smaller than the
raw data set.  We first multiply the data for each detector in each
TOD by a window function, reducing the weight in the first and last 10
seconds of each file, then remove a single offset and slope for the
entire 15-minute period.  There is no additional filtering in the time
domain, though the lowest frequencies are given no weight in the
process of maximizing the likelihood of the final map.  The detector
data are combined into a data vector $\vec{d}$.  The mapping is done
in a cylindrical equal-area projection with a standard latitude of
$\delta=-53.5\degree$ and pixels of $30\arcsec\times 30\arcsec$,
roughly one third of the beam FWHM\@. The map is represented as a
vector $\vec{m}$ of length $N_\mathrm{pix}\sim10^7$.  We model the
data as ${ \vec{d}=\mat{P} \vec{m}+\vec{n}}$ where the matrix
$\mat{P}$ projects the map into the time stream, and $\vec{n}$ is the
noise, which has covariance matrix $\mat{N}$. The maximum likelihood
solution, $\vec{\tilde m}$, is given by solving the mapping
equation~\citep[e.g.,][]{tegmark:1997a}:
\begin{equation}
{
\mat{P}^{\rm T} \mat{N}^{-1}\mat{P} \vec{\tilde m}=\mat{P}^{\rm T}\mat{N}^{-1}\vec{d}.
}
\label{eq:mapeq}
\end{equation}
We solve for $\vec{\tilde m}$ iteratively using a preconditioned
conjugate gradient (PCG) method
\citep{press/teukolsky/vetterling:NRC:3e, hinshaw/etal:2007}. 
Based on simulations, we find that the solution is an unbiased
estimator of the sky for $\ell>600$, the multipoles that we analyze
in this paper.  Additionally, the clear cross-correlation with WMAP 
(Section~\ref{sec:wmap_cal}) indicates that the maps are likely
unbiased as low as $\ell\approx 200$.  This approach
is different from that taken by other recent measurements of the fine
angular scale anisotropy in which the time stream data are filtered
and binned \citep[e.g.,][]{friedman/etal:2009, reichardt/etal:2009a,
lueker/etal:prep}.

Solving Equation~\ref{eq:mapeq} requires careful consideration of the
noise structure of the ACT data. First, the data are weighted in
Fourier space to account for the variation in noise with frequency,
particularly to reduce the noise that the atmosphere adds at low
frequencies.  For frequencies less than 0.25 Hz, we set the
statistical weights to zero.  Higher frequencies are given
successively more weight in inverse proportion to the noise variance
in each band.  Second, we find that there are several modes,
particular combinations of the 1000 detectors, that correspond to
signals unrelated to the celestial temperature; for example the
common mode, atmospheric gradients, and other detector correlations.  We
handle them by finding the ten modes with the largest eigenvalues
in each TOD and solving for their amplitudes as
a function of time along with the map $\vec{\tilde
m}$. The power spectrum (Section~\ref{sec:results}) is robust to
halving or doubling the number of modes.

Maps are made using only fractions of the valid data so that null
tests can be performed.  The subset maps are also used for finding the
CMB power spectrum, as described further in
Section~\ref{sec:spectrum_single}.  The final map used for
source-finding, however, is a complete run based on all the valid
data.  Figure~\ref{fig:maps} shows the map and the region used.  The
computational task is considerable. One iteration takes 100 seconds on
5000 cores of Canada's SciNet GPC cluster each running at 2.53\,GHz;
computations required before the first iteration take an additional
time equivalent to approximately twenty iterations. A converged map
requires hundreds of iterations and approximately ten CPU years.

\subsection{Calibration to WMAP}
\label{sec:wmap_cal}

The absolute calibration described in Section~\ref{sec:calibration} is
limited by the uncertainty in the \arone\ brightness of Uranus, the
primary calibrator.  We complement the
planetary calibration by cross-correlating with the W-band (94\,GHz)
measurements of WMAP \citep{hinshaw/etal:2009}.  In the angular scale
range of $200\le\ell\le1200$, both instruments measure the sky with
sufficient signal-to-noise to permit the comparison.

The technique requires a single map from WMAP and two maps with
independent noise from ACT\@.  We assume that only one relative
calibration ratio between ACT and WMAP must be estimated.  The WMAP
data set consists of the high-resolution\footnote{The WMAP maps are at
  HEALPix resolution $N_{\rm side}=1024$, with $3.5^\prime$ pixels.}
W-band five-year map.  The two independent ACT maps are each made with
one half the current data set.  We compare the maps only in the
\psSurveyArea\,deg$^2$ strip used for the present power spectrum
analysis (described in Section~\ref{sec:fields}), where the noise is
low and uniform.

We compute those cross-spectra in two-dimensional ${\ell}$-space
that combine either the full ACT data set with the WMAP map or the two ACT
maps with each other.  We average over the polar angle in ${\ell}$-space
to get a one-dimensional cross-spectrum for each.  In the case of the
ACT spectra, the noise is not isotropic, so the spectra are
angle-averaged with appropriate weights.  We then find the single
calibration factor that minimizes $\chi^2$ for a model in which the
minimum-variance weighted combination of the ACT-ACT and the ACT-WMAP
cross-spectra equals the WMAP all-sky spectrum.  The model accounts
for the fact that the ACT and WMAP measurements have noise varying as
different functions of $\ell$.

The result is a calibration factor with $<6\%$ fractional uncertainty
in temperature.  The calibration factors derived from WMAP and from
Uranus observations are consistent to $\lesssim 6\%$.

\section{FOREGROUNDS}
\label{sec:foregrounds}

Radio and infrared galaxies are the dominant sources of foreground
emission in the \aroneApprox\ band at $\ell \ge 1000$. To study the
underlying CMB spectrum, the sources must be identified and masked,
and residual contamination in the power spectrum must be accounted
for.  Many approaches to the problem have been described in the CMB
literature \citep[e.g.,] [] {wright/etal:2009, reichardt/etal:2009a,
reichardt/etal:2009, sharp/etal:prep, dawson/etal:2006,
sievers/etal:prep}. Our approach is to find the sources in the ACT
maps, mask them, and assess the residual contribution with models.

We find the sources using a matched filter with noise weighting
derived from the statistical properties of the map
\citep{tegmark/deoliveira-costa:1998}. We identify as sources all
pixels detected at $\ge 5\sigma$ having at least 3 neighboring pixels
detected at $\ge 3\sigma$.  A $5\sigma$ detection corresponds to
roughly 20\,mJy in the filtered map. The selection criteria are tested
on simulated sky maps to determine the sample's purity and
completeness. The sample is approximately $85\%$ complete at 20\,mJy
and $\sim100\%$ complete at 50\,mJy. For fluxes greater than 20\,mJy,
the detections have a purity of $\sim95$\% (that is, in a hundred
detections approximately five are false). Simulations show that at 20
mJy deboosting \citep{condon:1974} is a $\sim 9\%$ effect, and at 40
mJy a $\lesssim 1\%$ effect.

In the \psSurveyArea\,deg$^2$ area, we detect
\nmaskedSources sources. Of these, 105 can be identified with sources in the
PMN \citep{wright/etal:1994}, SUMSS \citep{mauch/etal:2003}, and/or
AT20G \citep{murphy/etal:prep} radio source catalogs. One source, not in
these catalogs, is in the 2MASS catalog (and thus has IR emission).
 Two have no previously measured counterparts.  In a $1\arcmin$
search radius, 17 sources have both radio and IR identifications. With
the exception of the single 2MASS source, all are treated as radio sources.

We fit the flux distribution of the radio sources to three models.
The distribution agrees with the \citet{toffolatti/etal:1998} radio
model scaled by $0.49\pm0.12$ in source counts, where the uncertainty
is statistical only. In the WMAP analysis of sources at
41\,GHz, \citet{hinshaw/etal:2007} found a good fit with the same
model after scaling the counts by 0.64. Thus it appears that on
average the radio population at \aroneApprox\ is composed
predominantly of flat-spectrum sources.  We prefer to compare the data
to models by scaling the model's flux rather than the counts, as the
extrapolation of the radio flux into the \aroneApprox\ range is
typically the most uncertain part of any model.  For
the \citet{toffolatti/etal:1998} model, the required flux scaling is
$0.5$.  The sources are also well fit by the
\citet{dezotti/etal:2005} model scaled by $1.2$ in flux. Lastly,
we fit the radio sources model in \citet{sehgal/etal:2010} and find
that its flux should be scaled by approximately $2.1$.  All scale
factors have a statistical uncertainty of about 25\%.  Each of
these models can be used to estimate the residual rms flux below the
20\,mJy cut, as discussed in Section~\ref{sec:other_ptsrc}.  We plan a
more thorough exploration of the radio sources in an upcoming
publication.

Before computing the power spectrum, we mask a $10\arcmin$ diameter
region around each of the \nmaskedSources sources. We call this
Mask-ACT.  In total, 2.3 deg$^2$ are masked,
or 1.0\% of the map.  Doubling the size of the masked
holes has a negligible effect on the power spectrum.  We also create a
cluster mask, Mask-C, by finding all $5\sigma$ clusters and check that
using this mask combined with Mask-ACT has a negligible effect on the
power spectrum. We also generate a mask, Mask-R, based on known radio
sources measured in the SUMSS (0.8\,GHz), PMN (5\,GHz) and ATCA
(20\,GHz) catalogs.  A model that has been fit to multiple data sets
\citep{sehgal/etal:2010} is used to estimate the mean spectral index,
and this index is then used to determine the flux cut in the radio
catalogs. Since the model is based on ensemble properties of radio
sources it may miss lower-flux low frequency sources
having shallow indices.  The resulting power spectrum is independent of which
radio source mask we use, or if we use the union of the two, and does
not depend on whether we include the cluster mask.

Diffuse dust emission becomes dominant near the galactic plane.  We
check for evidence of dust in the region of our map nearest to the
plane, from right ascension \hours{6} to \hours{7}.  Through a
cross-correlation with the estimated dust map of 
\citet{finkbeiner/davis/schlegel:1999} (the FDS map), we can limit the
dust contribution here to $\ell(\ell+1)C_\ell/(2\pi)\lesssim 5\,\uk^2$
at $\ell= 1000$, while a direct power spectrum of the FDS map in the
same region gives $\sim 1\,\uk^2$.  As the diffuse dust component
decreases with increasing $\ell$, it is not significant in our
analysis.

\section{POWER SPECTRUM METHOD}
\label{sec:method}

We estimate the CMB power spectrum using the adaptive multi-taper
method (AMTM) with prewhitening, described in
\citet{das/hajian/spergel:2009}.  We make independent maps from
subsets of the data and use only cross-spectra between maps to
estimate the final power spectrum.  All operations are performed using
the flat-sky approximation. 
We summarize the method in this section.

\subsection{Fields used for power spectrum analysis}
\label{sec:fields}

We find the power spectrum of our map by separate analysis of each of
the \npatch patches shown in Figure~\ref{fig:maps}. Each
patch is $4\fdeg2 \times 4\fdeg2$ in size, and together they cover a
rectangular area of the map from $\alpha =\hmin{0}{48}$ to
\hmin{6}{52} (12\degree\ to 103\degree) in right ascension and from
$\delta = \dm{-55}{16}$ to $\dm{-51}{05}$ in declination.  This area
is the region of a larger survey having the lowest noise.  We also
split the raw data into \ndivtext subsets of roughly equal size, with
the data distributed so that any \ndivtext successive nights go
into different subsets.  The \ndivtext independent maps generated from
these subsets cover the same area and have approximately the same
depth.  All maps are fully cross-linked.  That is, they all contain
data taken with the sky both rising and setting.

\subsection{Spectrum of a single patch}
\label{sec:spectrum_single}

We estimate the spectra of the \npatch patches independently, before
taking a weighted average to find the final spectrum.  There are
\ndivtext independent sky maps, from which \ncrosstext cross-spectra
and \nautotext auto-spectra are evaluated on each patch.  We use a
weighted mean of only the cross-spectra for the final spectral
estimate.  The weights depend on both the cross- and auto-spectra, as
discussed below.

Before separating the \ndivtext maps into \npatch patches, each
beam-convolved map, $T_b(\bm \theta)$, is initially filtered in
Fourier space with a high-pass  function $F_c(\ell)$.  This
filter suppresses modes at large scales that are largely
unconstrained. These modes arise from a combination of instrument
properties, scan strategy, and atmospheric contamination.
We choose a squared sine
filter, given in Fourier space by the smooth function
\be
\label{eq:highpass}
F_c(\ell) = \left\{
         \begin{array}
         {l@{\quad:\quad}l}
 0 & \ell <{\rm \ell_{min}} \nonumber \\
 \sin^2x(\ell) &{\rm \ell_{min}} < \ell < {\rm \ell_{max}}  \nonumber \\
 1 &  \ell < {\rm \ell_{max}}
 \end{array}
\right.  \ee 
where $x(\ell) \equiv(\pi/2)({\ell - {\ell}_{\rm min}})/({ \ell_{\rm
max} - \ell_{\rm min}})$. We choose $\ell_{\rm min}=100$ and
$\ell_{\rm max}=500$.  The \npatch patches of the map are treated
separately from this point.  The map of each patch is prewhitened
using a local, real-space operation to reduce the dynamic range of its
Fourier components \citep{das/hajian/spergel:2009}.  The prewhitening
operation involves adding a fraction (2\%) of the map to an approximation of
its Laplacian.  The Laplacian is computed in real space by taking the
difference between the map convolved with disks of radius 1\arcmin\
and 3\arcmin.
  Then the maps are multiplied by the point source mask.
The prewhitening step greatly reduces the leakage of power from low to
high multipoles caused by the action of the point source mask on the
highly colored CMB power spectrum.

For each patch, we compute the \ncrosstext 2D cross-spectra and
\nautotext 2D auto-spectra.  The axes correspond to right
ascension and declination.
Each spectrum is computed using the
adaptively weighted multi-taper method, using $N_{\rm tap}=5^2$ tapers
having resolution parameter $N_{\rm res} = 3$
\citep[see][]{das/hajian/spergel:2009}.  
Windowing the maps with 25 orthogonal taper functions allows us to
extract most of the statistical power available in the maps, at the
expense of broadening the resolution in angular frequency by a factor
of approximately $N_{\rm res}$.  At this stage, each 2D
cross-spectrum, $\tilde C_\bl^{i\alpha\beta}$, between submaps
$\alpha$ and $\beta$ on patch $i$ incorporates the effects of the
filter, prewhitening, tapering, the point source mask, and the beam; they are
analogous to a ``pseudo power spectrum'' of an apodized
map \citep[e.g.,][]{hivon/etal:2002}.  Each 2D spectrum is then
averaged in annuli with a narrow range of $|\bl|$ to give the binned
pseudo-spectrum ${\tilde {\cal B}_b}$, with
\be
{\tilde {\cal B}^{i\alpha\beta}_b} =P_{b \bl} \tilde {\cal
  B}^{i\alpha\beta}_\bl.
\ee
Throughout this paper we define ${\cal B}_\ell\equiv
\ell(\ell+1)C_\ell/2\pi$.  The binning function $P_{b \bl}$ is set to
one for pixels lying in an annular bin (indexed by $b$) of width
$\Delta\ell=300$ centered on $|\bl|=\ell_b$, and zero elsewhere.  The
size of the patches and the resolution of the tapers dictates the
width of the bins.  For our square patches of side $s=4\fdeg2$, the
fundamental frequency resolution in Fourier space is $\delta \ell =
2\pi/s \approx 90$.  The application of the tapers with a resolution
parameter $N_{\rm res}=3$ further degrades the resolution to $N_{\rm
res} \delta \ell$, so bins chosen to be smaller than
$\Delta\ell\sim270$ would be unavoidably correlated.  The binning function
is also set to zero where $|\ell_x|<270$.  This region of Fourier
space is particularly sensitive to scan-synchronous effects, either
fixed to the ground or in phase with the azimuth scan.

The binned pseudo spectrum $\tilde {\cal B}^{i \alpha\beta}_b$ is then
deconvolved with the mode-mode coupling matrix, which takes into
account the combined effects of tapering and masking and can be
computed exactly.  Lastly, we divide by the $\ell$-space representations
of the prewhitening filter, of the high-pass filter $F_c(\ell_b)$, and
of the beam
to obtain an unbiased estimate of the true underlying spectrum ${\cal
B}^{i \alpha\beta}_b$.

This procedure has been tested with simulations.  The number and
resolution of the tapers are chosen as the optimal balance between
maximizing information and minimizing bias caused by leakage of
power. The simulations confirm that increasing the number of tapers
beyond $5^2$ has a negligible effect on the spectrum errors.



\subsection{Combining patches}
\label{sec:combining_patches}

The final power spectrum estimator is given by a weighted mean over
$N=\npatch$ patches,
\be
\label{def:estimator}
{ \hat{\cal B}}_b = \frac{\sum_{i=1}^{N}  w_b^i  
{ \hat{\cal B}}_b^i }{\sum_{i=1}^{N} w_b^i},
\ee
where\footnote{Here we introduce the notation $\hat X$ as an unbiased
  estimator of the quantity $X$, in the sense $\ave{\hat X} = X .$}
  ${{ \hat{\cal B}}_b^i} \equiv \sum_{{\alpha,\beta; \alpha<\beta}} {
  {\cal B}}^{i\alpha\beta}_b /6$ is the mean of the \ncrosstext
  deconvolved cross-power spectra in patch $i$ (assuming equal
  weights), and $\alpha$ and $\beta$ index the four independent maps
  of that patch.  The weights are chosen as the inverse of the
  variance of this estimator in each patch, i.e.  $w^i_b =
  1/\sigma^2(\hat {\cal B}^i_b)$, where
\be
\sigma^2(\hat {\cal B}^i_b) \equiv 
\ave{ (\hat {\cal B}^i_b)^2 } - \ave{\hat {\cal B}^i_b}^2
\ee
and the average is taken over the several cross-spectra computed for
patch $i$.  The first term contains the 4-point function of the
temperature field and is approximated as,
\ba
 \nonumber\ave{ (\hat {\cal B}^i_b)^2 } &\simeq&  \frac2{M(M-1)}\\
 &\times&  \mathop{{\sum_{\alpha,\beta,\gamma,\delta}}}_{\alpha<\beta; 
\gamma<\delta} ({\cal B}^{\alpha\beta} _b {\cal B}^{\gamma\delta}_b
  +{\cal B}^{\alpha\delta}_b {\cal B}^{\beta\gamma}_b  + {\cal
    B}^{\alpha\gamma}_b {\cal B}^{\beta\delta}_b), 
\label{eq:temperature4point}
\ea
where $M=4$ is the number of submaps per patch. We have neglected
any connected (non-Gaussian) part of the 4-point function due to 
components such as point sources. This is a reasonable
approximation when choosing the \npatch weights, because the
expression (Equation \ref{eq:temperature4point}) is
dominated by the auto-spectrum terms, which in turn are
noise-dominated.

\subsection{Power spectrum covariance}
\label{sec:ps_uncertainties}

\begin{figure}[t]
  \centering
  \epsscale{1.28}
  \plotone{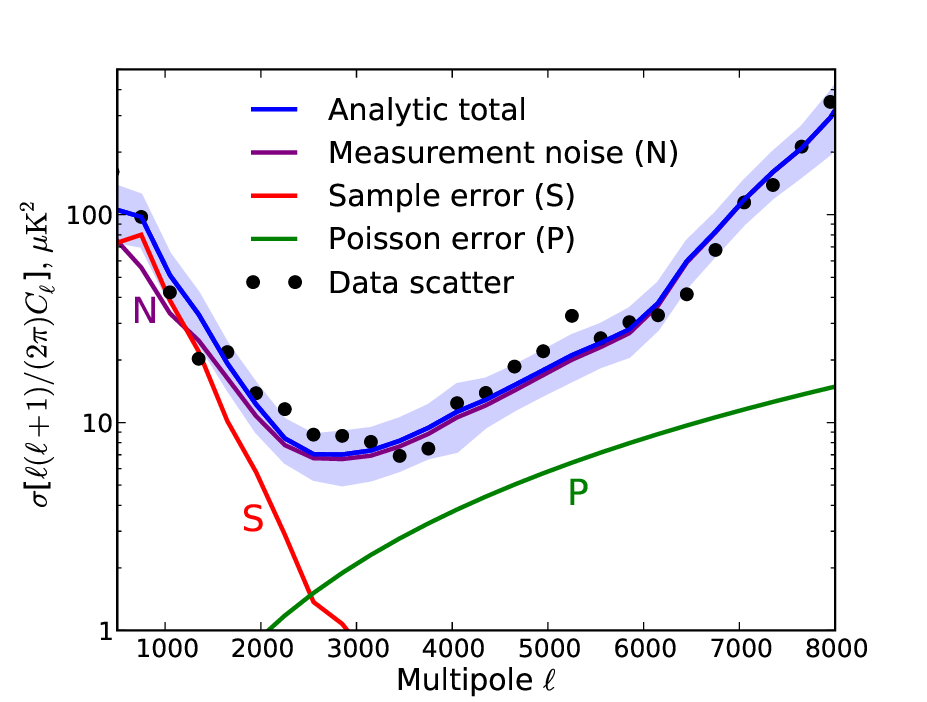}
  \caption{ The total estimated error $\sigma$ on the power spectrum (blue)
    given by the analytic expression
    (Equations~\ref{eq:analyticPSVariance} and
    \ref{eq:analyticPSPoissonTerm}).  The uncertainty on $\sigma$ is found
    using simulations and is shown by the shaded blue band.  The
    Gaussian sample variance (red line, labeled ``S'') dominates for
    $\ell\lesssim1200$, and atmospheric plus instrument noise (purple,
  ``N'') dominates at 
    $\ell>2000$. The non-Gaussian term due to unmasked point sources
    and clusters of galaxies contributes about 15\% of the variance at
    $2500<\ell<6000$ (green, ``P'').  The errors estimated using the
    scatter of the results from the \npatchtext patches (Equation
    \ref{eq:binBinCovariance} are shown for comparison (black points);
    they agree well with the analytic errors.  }

  \label{fig:analytic_errors}
\end{figure}

We estimate the bandpower covariance matrix $\mat\Sigma$ using the
scatter in the power spectrum among the $N=\npatch$ patches,
\begin{eqnarray}
{\Sigma_{bb'}} &\equiv& {\ave{\Delta {\cal B}_b \Delta {\cal B}_{b'}}} \\
&=&  \frac{\sum_{i=1}^{N} w_b^i
w_{b'}^i \ave{({\hat {\cal B}}_b^i -{\hat{\cal B}}_b)({\hat {\cal
B}}_{b'}^i- \hat{\cal B}_{b'})} }{\sum_{i=1}^{N} w_b^i \ \sum_{j=1}^{N}
w_{b'}^j}.
\label{eq:binBinCovariance}
\end{eqnarray}
The square roots of the diagonal elements of the covariance matrix are
reported as the errors on our power spectrum estimate.

To test the accuracy of the error estimate, the errors on the power
spectrum are also computed analytically. Three terms contribute:
sample variance in the CMB multipoles due to limited sky coverage,
instrumental and atmospheric noise, and a non-Gaussian term due to
unmasked point sources and galaxy clusters.  The diagonal component of
the variance in one patch can be written as the sum of these terms, in
order:
\be
\label{eq:analyticPSVariance}
\sigma^2( \hat C_b ) = 
          \frac{   2 {\hat C_b^2}}{n_b} + 
          \frac{ 4 \hat C_b \hat N_b/M + \hat N_b^2/n_w} {n_b}+
    { \frac{ \sigma^2_P}{f_{\rm sky}}},
\ee
where $\hat N_b = (\hat C_b^A - \hat C_b)$, the difference between the
weighted mean auto- and cross-spectra, estimates the average power spectrum
of the noise; $M=4$ is the number of maps with independent noise
properties per patch; $n_w= M(M-1)/2=6$ is the number of cross-spectra
per patch; $n_b$ counts the number of Fourier modes measured in bin
$b$ (that is, the number of pixels falling in the appropriate annulus
of Fourier space); $f_{\rm sky}$ is the patch area divided by the
full-sky solid angle, $4\pi$\,steradians; and $\hat C_b$ is the weighted
mean cross-spectrum.  In the last term, $\sigma^2_P$ is
given by the non-Gaussian  part of the four-point
function.  Thus
\be
 \label{eq:analyticPSPoissonTerm}
 \sigma^2_P=\frac{1}{4\pi}\left[\ave{\delta T^4(\hat{n})} - 3  \ave{\delta T^2(\hat{n})}^2\right],
\ee
where $T$ is the temperature map and the average is over pixels in the
map. The term arises from the
Poisson-distributed components in the maps, including unresolved point
sources and clusters of galaxies. For purposes of the covariance
calculation, we assume that such sources are distributed independently
of one other.  This term is constant as a function of $\ell$ and is
computed from the masked maps with the high-pass filter
(Equation~\ref{eq:highpass}) applied. The overall variance is given by the
weighted mean of the patch variances (Equation
\ref{eq:analyticPSVariance}). The covariance among bins is small in
the limit that the four-point term does not dominate, and with bins
chosen to be sufficiently large.

Figure \ref{fig:analytic_errors} shows the three components of the
errors.  The Gaussian part due to sample variance dominates below
$\ell=2000$; it scales as $C_b/\sqrt{n_b}$. The Poisson term due to
the non-Gaussian clusters and unresolved point sources contributes
about 15\% of the error budget for $2500\lesssim\ell\lesssim6000$ but
is sub-dominant at all scales. The noise term dominates at
$\ell\gtrsim2000$; atmospheric noise is the main contribution at
angular scales of $\ell\lesssim 1000$.  As a cross-check of the error
estimates, we also plot the errors derived from the scatter of the
power spectra from the \npatchtext\ patches.

The uncertainty in the analytic errors is shown as the shaded band in
Figure~\ref{fig:analytic_errors}.  It was estimated by Monte Carlo
simulations. One thousand patches were simulated with white noise and
their power spectra taken by the same methods used on the ACT maps.
The results demonstrate that the errors estimated from the scatter
among our
\npatch patches are consistent with the expected uncertainty.  The
same simulations were used to verify that the covariance between
different power spectrum bins (Equation~\ref{eq:binBinCovariance}) is
less than 1\%.


\section{POWER SPECTRUM RESULTS}
\label{sec:results}

\begin{figure*}[thb]
  \centering
\epsscale{1.0}
   \plotone{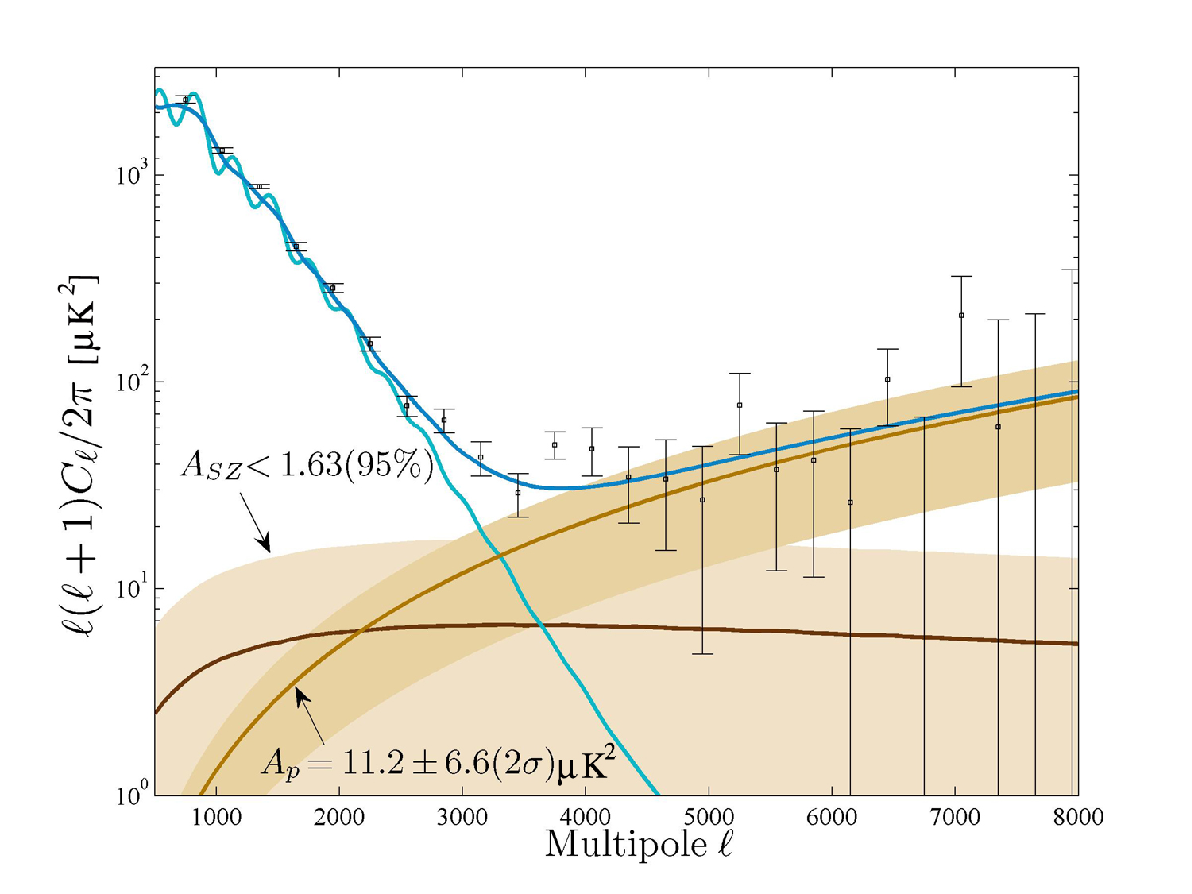}

   \caption{ The observed power spectrum in bandpowers at \aroneApprox\ from
     ACT observations (points with error bars). At large angular
     scales there is good agreement with the lensed \LCDM\
     model of the primary CMB (light blue curve shown for
     $\ell<4700$).  The $\chi^2$ of the model is 7.1 for 4 ACT data
     points in the range $600<\ell<1800$. The best-fitting model to
     the full dataset is shown (dark blue, the highest curve for all
     $\ell>2500$).  The complete model includes the primary CMB model
     plus both a Poisson power from point sources and SZ power
     from clusters; both additional components have been allowed to
     vary.  The complete model has been smoothed by convolution with a
     boxcar window function of width $\Delta\ell=300$; the
     primary CMB model has not been smoothed.
     The narrower, gold band shows the marginalized $95\%$~CL limits
     on the Poisson amplitude, while the curve indicates the best-fit
     amplitude $A_p = 11.9\,\uk^2$. The wider pink band shows the
     $95\%$ CL upper bound on the SZ amplitude, $A_{\rm SZ} < \maxASZ$;
     the dark curve inside it shows the best-fit value of $A_{\rm SZ}
     = \bestASZ$. The Poisson and SZ power are consistent with higher
     frequency observations and with $\Lambda$CDM predictions.  The
     fitting procedure is described in Section \ref{sec:params}.
   }

  \label{fig:cl_binned}
\end{figure*}

The binned estimate of the power spectrum $\hat{\cal B}_b$ is shown in
Figures \ref{fig:cl_binned} and \ref{fig:spectrum_many}, and
bandpowers are given in Table \ref{tab:cls}.  With our method the
bandpowers are estimated to have less than 1\% correlation with
neighboring bins, but the window functions have a small overlap, which
we account for in the analysis.  At $\ell\lesssim2500$, the estimated
power is consistent with previous observations by ground and
balloon-based experiments. The features of the acoustic peaks are not
distinguished with the coarse binning, but with the fluctuation
band-power measured to $5\%$, this spectrum offers a powerful probe of
cosmological fluctuations at small scales.  A clear excess of power is
seen at $\ell\gtrsim2500$ which can be attributed to
point sources and, to a lesser extent, to the SZ effect.

\begin{table}[b]
  \centering
  \caption{ \label{tab:cls}
    Anisotropy power in bands of width $\Delta \ell =300$.
  Temperatures are in CMB units.
  }
  \begin{tabular}{rr|rrrr}
 Central   & $\ell(\ell+1)C_b/2\pi$ & \LCDM &  SZ & Source &
 Total  \\
 $\ell_b$ & $(\uk^2$)\footnote{There is negligible covariance
 between bins.  The maximum-likelihood fit agrees best with \LCDM\ if
 the $C_b$ data given here are multiplied by 0.91.}\footnote{For
 comparison with the SPT results of \citet{lueker/etal:prep}, we also
 compute the spectrum with wider bins.  For five bins of width
  $\Delta\ell=400$ from $3000<\ell<5000$, we find $39\pm6$, $41\pm7$,
  $49\pm10$, $36\pm12$, and $35\pm13$.  For four bins of width
  $\Delta\ell=900$ from $5000<\ell<8600$, we find $56\pm19$,
  $45\pm25$, $140\pm 70$, and $80\pm250\,\uk^2$.  The ACT data are also shown in
  Figure~\ref{fig:spectrum_many} with bins of width $\Delta\ell=600$
  starting at $\ell=4200-4800$; 
the  bandpowers are given by $31\pm11$, $39\pm18$,
  $36\pm19$, $61\pm26$, $29\pm58$, and $28\pm116\,\uk^2$.} &
  model\footnote{This and all ``model'' columns are in the same
  units as the data:  thermodynamic  $\uk^2$.} &
 model\footnote{Assumes the best-fit value of $A_{\rm SZ}=\bestASZ$.}
 & model\footnote{Assumes the best-fit value of $A_p=\bestAP$ and
 uncorrelated Poisson sources.} & \\
\hline


 750 & $2317\pm 98$ & 2106.9 &  3.6 & 0.8 & 2111.2 \\
1050 & $1313\pm 42$ & 1204.2 &  4.6 & 1.5 & 1210.3 \\
1350 & $ 882\pm 20$ & 777.3 &  5.3 & 2.4 & 785.0 \\
1650 & $ 450\pm 22$ & 431.9 &  5.8 & 3.6 & 441.3 \\
1950 & $ 284\pm 14$ & 248.3 &  6.1 & 5.0 & 259.5 \\
2250 & $ 153\pm 12$ & 133.8 &  6.3 & 6.7 & 146.8 \\
2550 & $  77\pm  9$ & 72.3 &  6.5 & 8.6 & 87.4 \\
2850 & $  65\pm  9$ & 38.0 &  6.6 & 10.8 & 55.3 \\
3150 & $  43\pm  8$ & 19.8 &  6.7 & 13.1 & 39.6 \\
3450 & $  29\pm  7$ & 10.3 &  6.7 & 15.8 & 32.7 \\
3750 & $  50\pm  7$ & 5.3 &  6.7 & 18.6 & 30.6 \\
4050 & $  47\pm 12$ & 2.9 &  6.6 & 21.7 & 31.2 \\
4350 & $  35\pm 14$ & 1.6 &  6.6 & 25.0 & 33.2 \\
4650 & $  34\pm 19$ & 0.9 &  6.5 & 28.6 & 36.0 \\
4950 & $  27\pm 22$ & 0.6 &  6.4 & 32.4 & 39.4 \\
5250 & $  77\pm 33$ & 0.4 &  6.3 & 36.5 & 43.1 \\
5550 & $  38\pm 25$ & 0.3 &  6.2 & 40.7 & 47.3 \\
5850 & $  42\pm 30$ & 0.2 &  6.1 & 45.3 & 51.6 \\
6150 & $  26\pm 33$ & 0.2 &  6.0 & 50.0 & 56.2 \\
6450 & $ 103\pm 41$ & 0.1 &  5.9 & 55.0 & 61.1 \\
6750 & $ -30\pm 68$ & 0.1 &  5.8 & 60.3 & 66.2 \\
7050 & $ 209\pm115$ & 0.1 &  5.7 & 65.7 & 71.6 \\
7350 & $  60\pm139$ & 0.1 &  5.6 & 71.4 & 77.1 \\
7650 & $ -38\pm213$ & 0.1 &  5.5 & 77.4 & 83.0 \\
7950 & $ -63\pm349$ & 0.1 &  5.4 & 83.6 & 89.1 \\


\end{tabular}
\end{table}

\begin{figure}[htb]
  \centering 
  \epsscale{1.2} 
  \plotone{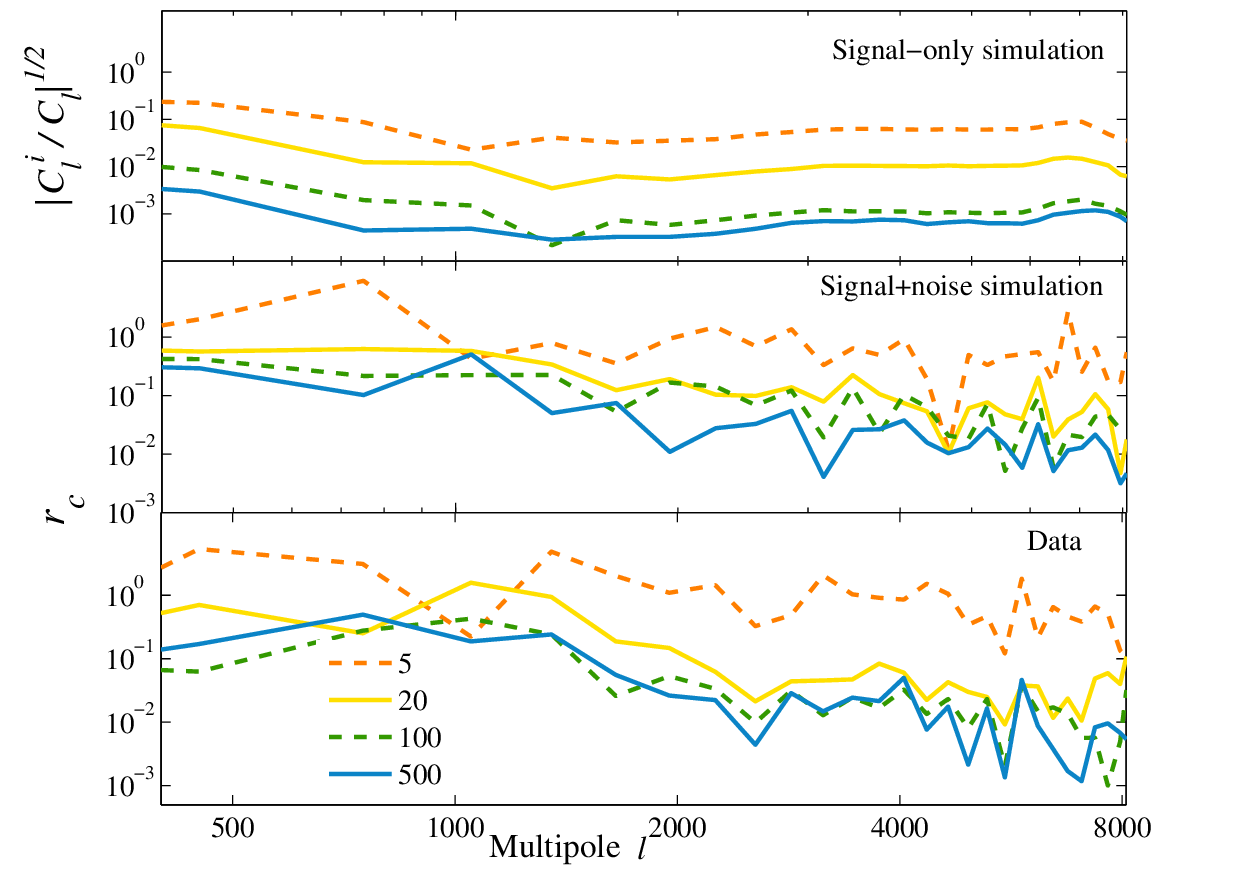} 

  \caption{Convergence of the maps as a function of iteration. The
    mapmaking algorithm converges well by iteration 500. For
    signal-only simulations (\emph{top}), the amplitude of
    fluctuations in the difference between the processed output map at
    iteration $i$ and the input map (denoted $({C^i_b})^{1/2}$), is
    less than 1\% of the amplitude of fluctuations in the input map
    ($\sqrt{C_b}$) at all scales by $i=500$. Iterations $i=5,20,100,$
    and 500 are shown. For simulations with noise and for the data
    (\emph{middle and bottom}, respectively), convergence is tested by
    estimating the maximum change in power between the processed map
    at iteration $i$ and iteration $1000$, as a fraction of the
    uncertainty in the power in the final map. For iteration 500 this
    fraction $r_c$ (described in Section \ref{sec:results}) is
    sufficiently small, less than 0.5 at all scales.  }
  
  \label{fig:diff_map_spectra}
\end{figure}

The large-scale modes are recovered only after iterating the maps.  To
test for convergence of noiseless maps, we compute the spectra ${\hat
  C}^i_b$ of difference maps, between the processed simulation map and
input simulated sky, at successive iteration numbers $i$. To account
for differences that are non-uniform between data subsets, we use the
auto-spectra. We show that the amplitude of fluctuations in the
difference map ($\sqrt{{\hat C}^i_b}$) is small, less than 1\% of the
amplitude of fluctuations in the input map ($\sqrt{{\hat C}_b}$) at
all scales by iteration 500, as shown in Figure
\ref{fig:diff_map_spectra}.  To test for convergence in the data,
where we do not know the input map, we estimate
the maximum change in power between the processed map at iteration $i$
and the final iteration, estimated as $2\sqrt{{\hat C}^i_b {\hat
    C}_b}$ using auto-spectra. Here ${\hat C}^i_b$ is the spectrum of
the difference map between iteration $i$ and the final iteration,
number 1000. We define the convergence ratio $r_c$ as this change in
power given as a fraction of the uncertainty in the power, $\sigma({\hat
  C}_b)$, and find it to be sufficiently small (less than 0.5)
by iteration 500 at all scales. The cross-correlation with WMAP
suggests that the maps are well converged down to $\ell\sim200$.  We
do not divide the angular power spectrum by a transfer function at any
value of $\ell$.

\begin{figure}[t]
  \centering
  \epsscale{1.3}
  \plotone{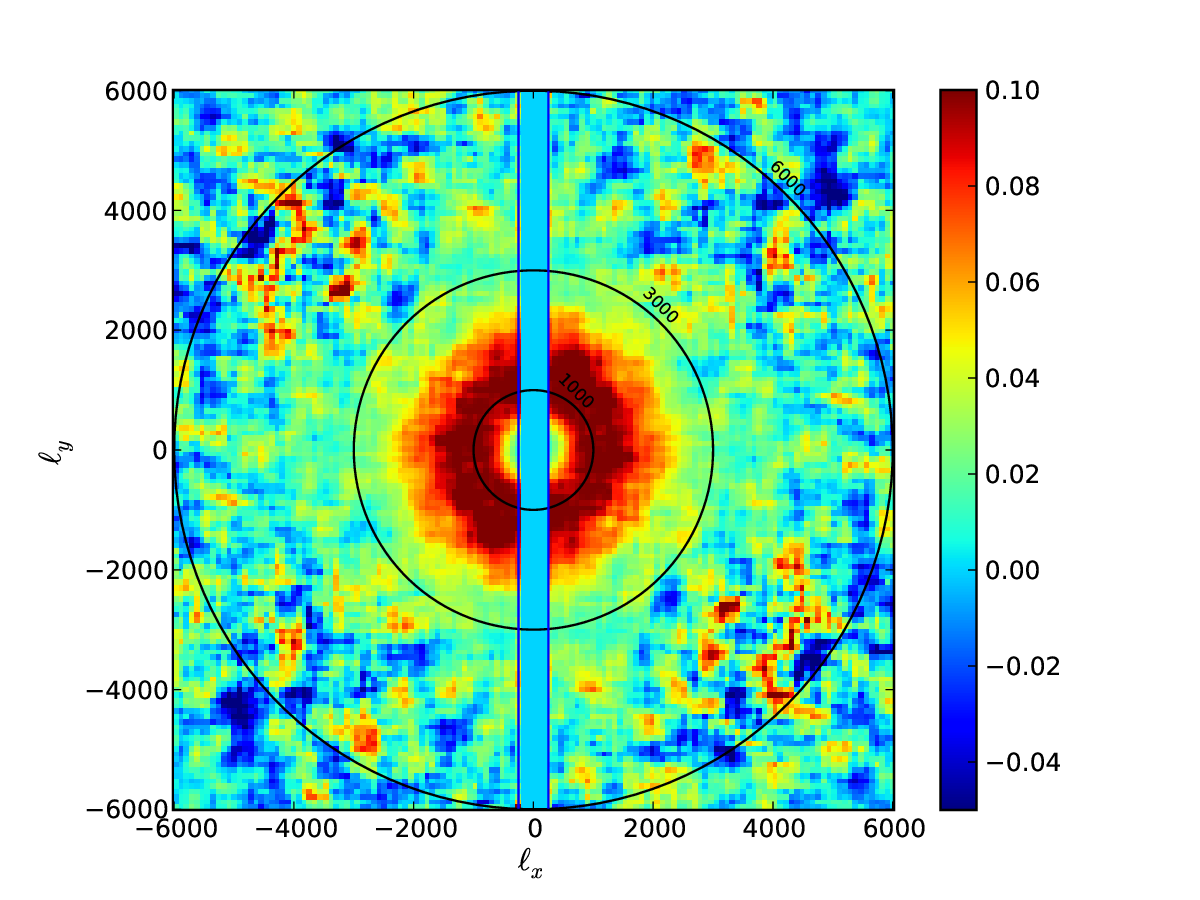}

  \caption{ The estimated two-dimensional power spectrum ${ C}_\bl$,
    multiplied by a factor of $\ell$ to emphasize the angular, rather
    than the radial, variation.  The power is consistent with being
    isotropic, when divided into wedges of $\Delta
    \theta=20\degr$. The vertical lines indicate the narrow region
    $|\ell_x|<270$, where excess power from scan-synchronous signals
    contaminates the power spectrum.  This region is not used for the
    power spectrum analysis.  The regions near
    $(|\ell_x|,|\ell_y|)\approx(4000,4000)$ are more noisy but not
    biased.  A power spectrum computed without these regions is
    consistent with the one we present.  }
  \label{fig:isotropy}
\end{figure}

We test the isotropy of the power spectrum by estimating the power as
a function of phase $\theta = \tan^{-1} (\ell_y/\ell_x)$. We compute
the inverse-noise-weighted two-dimensional pseudo spectrum coadded over
map subsets and patches.  The mean cross-power pseudo spectrum is
shown in Figure \ref{fig:isotropy}, indicating the region masked at
$|\ell_x|<270$. The spectrum is symmetric for $\bl$ to $-\bl$, as it
is for any real-valued maps.  To quantify any anisotropy, the
power averaged over all multipoles in the range $500<\ell<8000$ is
computed in wedges of $\Delta \theta = 20 \degr $. It is found to be
consistent with an isotropic 2D spectrum.

\begin{figure}[t]
  \centering \epsscale{1.15} 
  \plotone{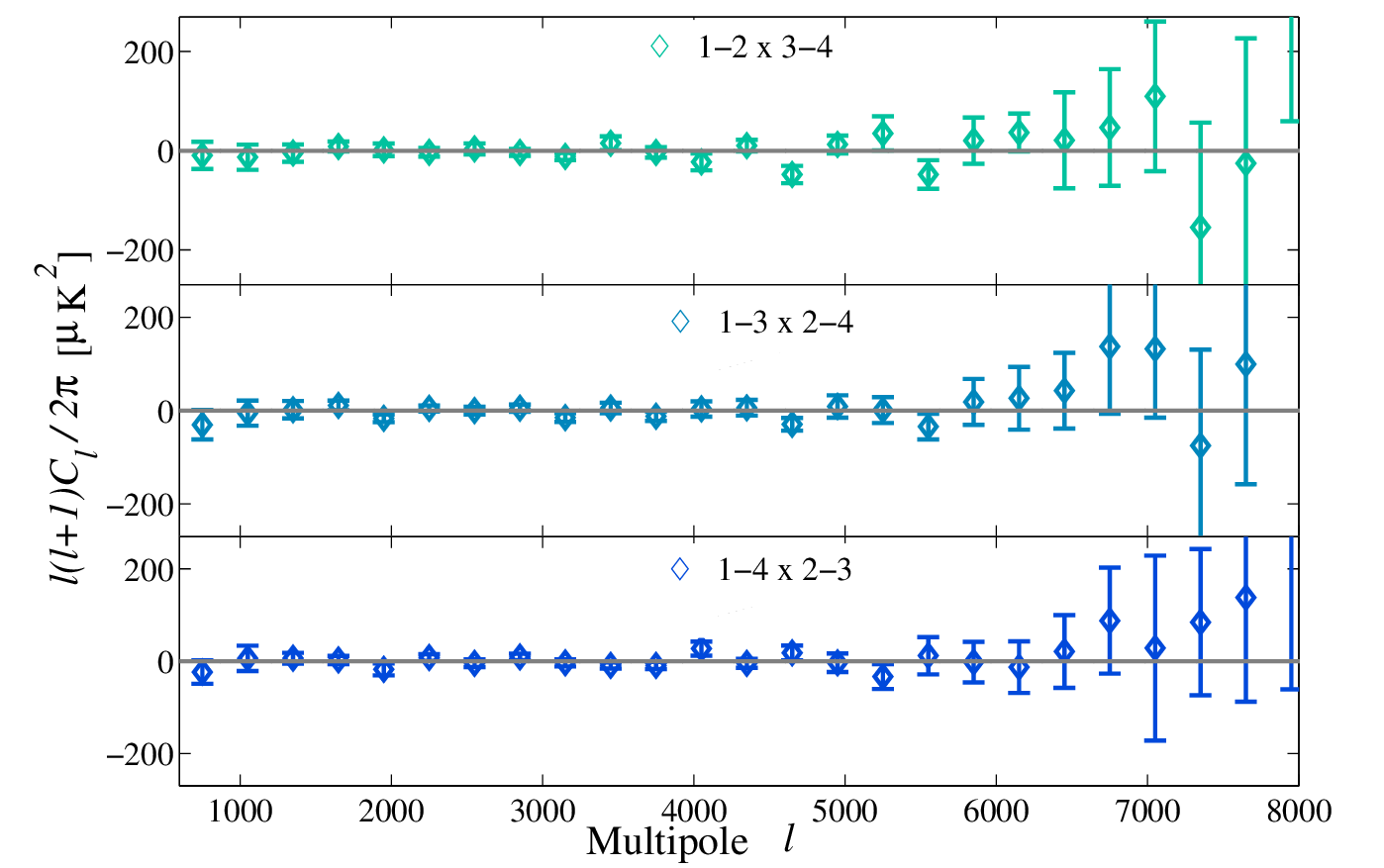}
  \caption{ Cross-spectra of difference maps formed from four data
    subsets. In the top panel, two maps $T^{12}\equiv T^1-T^2$ and
    $T^{34}\equiv T^3-T^4$ are formed; both are expected not to contain
    signal. The cross-spectrum is consistent with no signal. The lower
    two panels show the same cross-spectra for the other two
    permutations of the four data subsets and are also consistent with
    no signal ($\chi^2$ is 25.2, 27.1, and 28.4 in the three panels
    with 25 degrees of freedom).}
  \label{fig:jackknife}
\end{figure}

We test that the signals in separate data subsets are consistent 
by taking the cross-spectrum of two difference maps formed from 
the temperature maps, $T^i$, of the four data subsets via:
\ba
T^{12}(\hat n) &\equiv& [T^1(\hat n)-T^2(\hat n)]/2\\
T^{34}(\hat n) &\equiv& [T^3(\hat n)-T^4(\hat n)]/2.
\ea
(The data subsets are described in Section \ref{sec:fields}.)  The
difference maps are expected to contain noise but no residual signal.
We estimate the cross-spectrum of the difference maps, 
${\hat C}_b = \ave{\tilde T^{12} \tilde T^{34}}$ using the 
methods described in Section \ref{sec:method} for $M=2$ data segments. 
The two other permutations of 
the data, ${\hat C}_b = \ave{\tilde T^{13} \tilde T^{24}}$, and 
$\ave{\tilde T^{14} \tilde T^{23}}$ are also tested. 
The three difference spectra, shown in Figure \ref{fig:jackknife}, are 
consistent with no signal.
%

\section{CONSTRAINTS ON SZ AND IR EMISSION}
\label{sec:params}

\begin{figure*}[t]
  \centering
  \includegraphics[width = 1\textwidth]{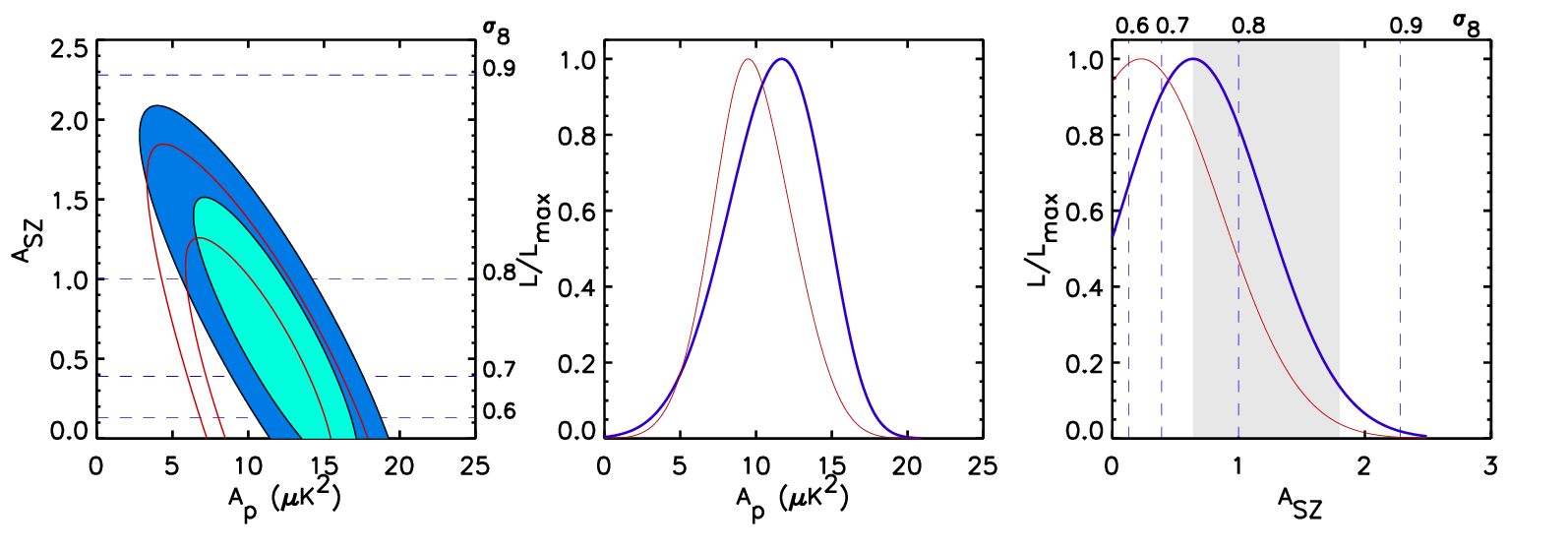}

  \caption{Probability distributions for the amplitude $A_p$ of an IR
  Poisson point source contribution of the form ${\cal
  B}_\ell\equiv\ell(\ell+1)C_\ell/2\pi = A_p(\ell/3000)^2$, and of
  $A_{\rm SZ}$, the amplitude of the SZ effect relative to that of a
  model with $\sigma_8=0.80$.  Each panel shows two models for the
  point source power spectrum.  In one case, we assume the point
  sources are uncorrelated Poisson-distributed sources.  In the second
  case, we marginalize over the amplitude of a correlated term scaling
  as ${\cal B}_\ell \propto \ell$.
  \emph{Left:} The two-dimensional distribution of $A_p$ and $A_{\rm
  SZ}$.  The filled (blue) regions assume uncorrelated sources; the
  unfilled (red) regions allow for correlated sources.
  \emph{Center:} The distribution of $A_p$ marginalized over $A_{\rm
  SZ}$.  In the center and right panels, the curve peaking at higher
  amplitude (blue) assumes uncorrelated sources. The point
  source power is consistent with SCUBA and BLAST data at higher
  frequencies. 
  \emph{Right:} The distribution of $A_{\rm SZ}$ marginalized over
  $A_p$.  The shaded region shows the 95\% CL limits on $\sigma_8$
  inferred from WMAP data combined with distance priors. }
\label{fig:data_ir_sz_amp}
\end{figure*}

We perform a simple analysis of the power spectrum to quantify the
combined contribution from dusty galaxies and radio sources, and the
level of SZ emission. We assume a \LCDM\ cosmology with lensing of the
CMB and parameters from the five-year WMAP analysis combined with BAO
and supernovae measurements \citep{komatsu/etal:2009}. We defer a full
investigation of cosmological parameter constraints until we improve
the absolute calibration and better account for astrophysical
foregrounds.

Our model for the power at \arone\ is 
\be
\label{eq:model}
{\cal B}_\ell^{\rm{th}} = {\cal B}_\ell^{\rm{CMB}}
+ A_{\rm SZ}{\cal B}_\ell^{\rm{SZ}} 
+ A_p\left(\frac{\ell}{3000}\right)^2 
+ {\cal B}_\ell^{\rm corr},
\ee
where ${\cal B}_\ell^{\rm {CMB}}$ is the lensed primary CMB power
spectrum; ${\cal B}_\ell^{\rm SZ}$ is a template spectrum
corresponding to a prediction for the SZ emission in a model with
$\sigma_8=0.8$ at \arone; $A_p$ quantifies the Poisson point source
power, required to be positive; and ${\cal B}_\ell^{\rm corr}$
corresponds to correlated point source power from clustered galaxies.
The SZ template we use includes the correlated thermal and kinetic SZ
(kSZ) effect derived from numerical simulations and is described in
detail in \citet{sehgal/etal:2010}. Its amplitude is assumed to scale
with $\sigma_8$ as the seventh power, such that $A_{\rm SZ} =
(\sigma_8^{\rm SZ}/0.8)^7$ for fixed baryon density. This accounts
approximately for the frequency-dependent combination of the thermal
SZ component scaling as the 7.5 to 8th power of $\sigma_8$ and the
sub-dominant kSZ scaling as the fifth power.  The expected point source
power $C_p$ is given by
\be
C_p = g(\nu)^2 \int_0^{S_{\rm cut}} S^2 \frac{\dif N}{\dif S} \dif S\ \ (\uk^2\,{\rm sr}),
\ee
an integral over all sources with flux $S$ up to some maximum flux
$S_{\rm cut}$, where $g(\nu) \equiv (c^2/2k\nu^2) \times [({\rm
  e}^x-1)^2/x^2{\rm e}^x]$ converts flux density to thermodynamic
temperature, and $x\equiv h\nu/kT_{\rm CMB}$.  Then $A_p$ is
the binned $\ell(\ell+1) C_p /2\pi$ Poisson power at pivot $\ell_0
=3000$. In thermodynamic $\uk^2$, $C_p=0.698 \times 10^{-6} A_p$. The
conversion to \jysr\ at \arone\ is $C_p [{\jysr}]= 1.55 C_p [10^{-5}
\uk^2\,{\rm sr}]$.

The infrared sources are expected to be clustered.  At small angles
galaxies cluster with typical correlation function $C(\theta)\propto
\theta^{-0.8}$ \citep[e.g.,][]{peebles:1980}, which would give $C_\ell
\propto \ell^{-1.2}$ on non-linear scales. Motivated by this, we first
adopt a simple
template for the correlated power,
\be
{\cal  B}_\ell^{\rm corr} = A_c \left(\frac{\ell}{3000}\right),
\ee
and fit for the amplitude $A_c$. Note that on larger scales,
$\ell<300$, ${\cal B}_\ell^{\rm corr}$ is expected to flatten and
gradually turn over \citep[e.g.,][]{scott/white:1999}, but at these
scales at \arone\ the CMB dominates. Models suggest that the power
from source clustering is less than the Poisson component at scales
smaller than $\ell=2000$ \citep{scott/white:1999, negrello/etal:2007,
righi/hernandez-monteagudo/sunyaev:2008}.
This is consistent with observations at 600\,GHz by BLAST
\citep{devlin/etal:2009, viero/etal:prep}. We therefore impose a prior
that the correlated power be less than the Poisson power at
$\ell=3000$, i.e., $0<A_c < A_p$.  This model is likely too
simplistic, and the correlated power may have an alternative shape
(e.g., \citealt{sehgal/etal:2010}, or the halo model considered in
\citealt{viero/etal:prep}). If the power from dusty galaxies is
instead better described by a linear matter power spectrum scaled by a
bias factor, ${\cal B}_\ell^{\rm corr}$ becomes almost degenerate with
the SZ component ${\cal B}_\ell^{\rm SZ}$ (see, e.g.,
\citealt{hall/etal:prep}).  Since it is not possible to separate these
components with data at a single frequency, our limit on the SZ
contribution at \arone\ should be considered as the upper limit on the
sum of the SZ power and a degenerate correlated source
power. Multi-frequency data will enable us to investigate the shape
and amplitude of the correlated component more fully.

Given this model, the likelihood of the data is given by
\be
 -2 \ln\mathscr{L} = ({\cal B}_b^{\rm th} - {\cal B}_b)^{\rm T} 
 \mat{\Sigma}^{-1}({\cal B}_b^{\rm th} -
  {\cal B}_b) + \ln \det\mat{\Sigma},
\ee
with covariance matrix $\mat{\Sigma}$ defined in
Equation~\ref{eq:binBinCovariance}.  The theoretical spectrum
$B_b^{\rm th}$ is computed from the model using ${\cal B}_b^{\rm th} =
w_{b\ell}\cal{B}_\ell^{\rm th}$, where $w_{b\ell}$ is an approximate
form of the bandpower window function in band $b$.  We marginalize
over the calibration uncertainty analytically
\citep{ganga/ratra/sugiyama:1996,bridle/etal:2002}. 

The uncertainty on the shape of the window function is small, of order
1.5\% for the \arone\ band, compared to 12\% overall calibration
uncertainty in power.  The window uncertainty is therefore neglected
throughout this analysis.  We verify that this approximation is valid
by including the window function uncertainty in the likelihood
calculation.  The beam Legendre transform is first expanded in
orthogonal basis functions, and the uncertainties on the basis
function coefficients are used to derive the window function
covariance matrix.  The covariance is dominated by a small number of
modes, so a singular value decomposition is taken.  The ten largest
modes are included in the likelihood, following the method described
in Appendix~A of \cite{hinshaw/etal:2007}.  We find that including the
window function uncertainty has only a negligible effect on parameter
estimates ($<0.05\sigma$).  The ACT/MBAC beam measurements and the
orthogonal function expansion are described in detail
in \citet{hincks/etal:prep}.

We confirm that the expected values of parameters $A_{\rm SZ}$, $A_p$,
and $A_c$ are recovered from maps of the ACT simulated sky.  We tested
simulations with and without noise in the map, and simulations with
realistic timestream noise run through the mapper.

\begin{table*}[tb]
  \centering
  \caption{ \label{tab:power_spectrum}
    Constraints on secondary anisotropies and extra-galactic
  foreground emission at \aroneApprox.}

\ifnum\manuscript=1
\begin{sideways}
\fi
  \begin{tabular}{c|cc|ccc|c}
    & $A_{\rm SZ}$ & $\sigma_8^{\rm SZ}$ & $A_p (\ell=3000)$ & $C_p$ &
$C_p$ & $\chi^2$/dof\\
    &  &  &  $\uk^2$& $10^{-5} \uk^2$  & ${\rm Jy}^2\,{\rm
sr}^{-1}$ & \\
\hline
  Poisson point sources    & $<1.63$ & $ <\maxSigmaEight$ &  $11.2\pm 3.3 $ & $0.78\pm0.23$   & $1.22\pm0.36$ &27.0/23\\
Poisson + correlated point sources    & $<1.36$ &  $<0.84$ &  $9.7\pm2.8$ & $0.68\pm0.20$   & $1.05\pm0.30$ &26.7/22\\
\hline
     \end{tabular}
\ifnum\manuscript=1
\end{sideways}
\fi
\vskip 1.cm
\label{table:params}
\end{table*}


In the range $600 < \ell< 1800$, where foreground emission and
secondary effects are sub-dominant, the data are consistent with the
lensed \LCDM\ model alone, with $\chi^2$ = 7.1 for four degrees of
freedom (dof).  If we rescale the maps to check consistency with the
model (multiplying temperatures by 0.96, a $0.7\sigma$ change in the
calibration), then the $\chi^2$ value becomes 3.0 with 3 dof.  We do
not rescale the maps in the analysis that follows, though the scale
factor is allowed to vary when we marginalized over uncertainties.

Using the full range $600<\ell<8100$, we find marginalized constraints
on {{$A_p= \meanAP\pm3.3\,\uk^2$ (thus, $C_p = (\bestCP\pm0.23$)
    \cpuK) and $A_{\rm SZ} <\maxASZ$ (95\% CL)}}. The minimum $\chi^2
= 27.0$ for 23 dof.  Assuming the scaling of $A_{\rm SZ}$ as
$(\sigma_8^{SZ})^7$, this implies an upper limit of $\sigma_8^{\rm SZ}
< \maxSigmaEight$ (95\% CL).  The one- and two-dimensional distributions are
shown in Figure~\ref{fig:data_ir_sz_amp}, with limits given in Table
\ref{table:params}. Marginalizing over the possible SZ power, ACT
detects a residual point source component at $3\sigma$ ($\delta \chi^2
=10$).  Marginalizing over a correlated term with $A_c<A_p$ gives
$A_p= 9.7\pm2.8$ $\uk^2$, and $\sigma_8^{\rm SZ} < 0.84$ with $\chi^2
= 26.7$ for 22 dof. Because $A_c$ is forced to be positive, its
inclusion has the effect of lowering the limits on $A_p$ and $A_{\rm
  SZ}$. Nevertheless, we do not find evidence for a correlated
component with current sensitivity levels.  The estimated parameters
vary by less than $0.6 \sigma$ when the minimum angular scale is
varied in the range $5000<\ell_{{\rm max}}<8000$, or the SZ template
is replaced by the spectrum of \citet{komatsu/seljak:2002}, which is
approximately 15\% lower than our template in the relevant range of
$\ell$ from 1000--5000.  We note that there are multiple models for
predicting $A_{\rm SZ}$ (e.g., \citealt{bond/etal:2005,
  kravtsov/nagai/vikhlinin:2005}) and that the relation between
$A_{\rm SZ}$ and $\sigma_8$ is an active area of research.



We also combine the ACT spectrum with the WMAP 5-year data
\citep{dunkley/etal:2009} to 
constrain the six-parameter \LCDM\ model (defined by the baryon
density, cold dark matter density, cosmological constant, optical
depth to reionization, and the amplitude and scale dependence of
primordial fluctuations at $k=0.002$ Mpc$^{-1}$). We model the SZ and
point source contribution using Equation~\ref{eq:model}, neglecting a
correlated component. We find similar results for the point source and
SZ amplitude in this extended model, with $A_p = 11.5 \pm 3.2$, and
$A_{\rm SZ} < 1.66$ (95\% CL). The \LCDM\ parameters are consistent with
WMAP alone, with $100\Omega_bh^2 =2.27\pm0.06$, $\Omega_ch^2 =
0.111\pm0.006$, $\Omega_\Lambda = 0.738\pm0.030$, $n_s=0.964\pm0.014$,
$\tau=0.086\pm0.017$, and $10^9 A_s = 2.4 \pm 0.1$.

\subsection{Comparison to other point source observations} 
\label{sec:other_ptsrc}

The residual source level, $C_p = (\bestCP \pm 0.23)$ \cpuK, combines
power from radio sources and dusty galaxies that were not removed
by the mask.  For the three radio source models discussed in Section
\ref{sec:foregrounds}, we expect residual power of $A_p = 6.4\,\uk^2$
(\citealt{toffolatti/etal:1998}, after rescaling by 0.4), $A_p =
4.1\,\uk^2$ \citep{dezotti/etal:2005}, and $A_p = 7\,\uk^2$ 
\citep{sehgal/etal:2010}. These models' amplitudes correspond to
$C_p^\mathrm{Radio} = 0.43, 0.29, {\rm and}\ 0.49$\,\cpuK,
respectively. The correction for those few sources below 20\,mJy that
we mask can be neglected; it is smaller than the spread among the
models.  Given the uncertainties in the models, we subtract the
typical model from the total source level and infer that the component
from residual IR sources lies in the approximate range $0.2 \lesssim
C_p^\mathrm{IR}\times10^5 \lesssim 1\,\uk^2$.

The ACT result on total point source power is similar to those of APEX
 at 150\,GHz, which finds $C_p =
1.1^{+0.9}_{-0.8}$ \cpuK\ \citep{reichardt/etal:2009a}, of ACBAR
($C_p = 2.7^{+1.1}_{-2.6}$ \cpuK,
\citealt{reichardt/etal:2009}), and SPT ($C_p =
(0.74\pm0.06)\cpuK$, \citealt{hall/etal:prep}).  This last measurement
employs a lower flux level for removing discrete sources:
6.4\,mJy versus our cut at 20\,mJy.  While this difference means that
the $C_p$ presented will contain roughly three times as much  power from
radio sources as the SPT measurement, we nevertheless find that the
total residual power due to point sources is consistent between the
two results, given the conservative assumption that at least one-quarter
of the point source power observed by ACT is due to dusty galaxies
rather than to radio sources.

The IR source models in the literature for 148\,GHz make a range of
predictions.  For example, for sources less than 20\,mJy
the \citet{lagache/etal:2004} model gives $C_p = 40$ \cpuK,
the \citet{negrello/etal:2007} model gives $C_p = 1.2$ \cpuK, and
the \citet{sehgal/etal:2010} model gives $C_p = 17.5$ \cpuK. The best
agreement comes with the \citet{negrello/etal:2007} model.

The emission from dusty galaxies in the rest frame can be modeled as
$S_0(\nu) \propto \nu^\beta B_\nu(T)$, with emissivity index $\beta
\sim 1.5$ and Planck function $B_\nu(T)$. The effective index
$\alpha$, where $S(\nu) \propto \nu^{\alpha}$ ($C_\ell \propto
\nu^{2\alpha}$), accounts for the redshift of the sources, the
intrinsic temperature variation $T$, and the index $\beta$. SCUBA has
observed emission at 850 \micro\meter\ (353\,GHz), where we expect a
similar population of galaxies to contribute
\citep[e.g.,][]{greve/etal:2004, greve/etal:2008}.  Using SCUBA galaxy
number counts and a model for $\dif N/\dif S$,
\citet{scott/white:1999} estimate the Poisson power to be $C_p =
730$\cpuK ($190$ \jysr) for $S_{\rm cut}=50$\,mJy.  Combined with
residual IR source level observed by ACT, this implies an effective
spectral index of $\alpha_{150-350}$ between 2.6--3.3. This is
consistent with emission from dusty star-burst galaxies at high
redshift and in line with predictions by \citet{white/majumdar:2003}
and \citet{negrello/etal:2007}.  It also agrees with the
$\alpha=2.6\pm0.6$ index inferred from source fluxes measured with
MAMBO (1.2\,mm) and SCUBA (850\,\micro\meter) \citep{greve/etal:2004},
  and with $\alpha=2.3$ measured from AZTEC (1.1\,mm) and SCUBA
  \citep{chapin/etal:prep}.  Observations by BLAST at 500
  \micro\meter\ (600\,GHz) have Poisson power $(2.7 \pm0.2) \times
  10^3$ \jysr\ \citep{viero/etal:prep}. Combining this with the ACT
  data leads to an estimate of the effective index
  $2.7\lesssim\alpha_{150-600}\lesssim3.6$, consistent with findings
  by APEX. The consistency with $\alpha_{150-350}$ suggests that
  similar populations are being probed at these frequencies, although
  BLAST is sensitive to a lower redshift range than ACT.

\subsection{Comparison to other SZ observations}

The ACT constraints on the SZ power indicate an amplitude of
fluctuations $\sigma_8^{\rm SZ} < \maxSigmaEight$ (95\% CL).  This
result is consistent with estimates that combine the primordial CMB
anisotropy with distance measures, $\sigma_8 = 0.81\pm0.03$
\citep{komatsu/etal:2009}, and improves on SZ-inferred limits at
150\,GHz from Bolocam ($\sigma_8^{\rm SZ}<1.57$)
\citep{sayers/etal:2009}, Boomerang ($\sigma_8^{\rm SZ}<1.14$ at 95\% CL)
\citep{veneziani/etal:2009},
and APEX ($\sigma_8^{\rm SZ}<1.18$)
\citep{reichardt/etal:2009a}.  We do not see evidence for an excess of
SZ power, in contrast to lower frequency observations by CBI at
30\,GHz which prefer a value $2.5\sigma$ higher than the concordance
value ($\sigma_8^{\rm SZ} = 0.922\pm0.047$,
\citealt{sievers/etal:prep}).

The ACT result is also consistent with the recently reported
$\sigma_8^{\rm SZ} = 0.773\pm0.025 $ ($A_{\rm SZ} = 0.42\pm0.21$) from
the South Pole Telescope \citep{lueker/etal:prep}. The SPT team prefers a
form for the correlated point sources that is covariant with the SZ
template in the $\ell=3000$ range \citep{hall/etal:prep}.  If such a
form is correct, then the $A_{\rm SZ}$ we report should be interpreted
as an upper limit on correlated point sources plus the SZ effect.
Analysis of the ACT's 218 and 277 GHz data will shed light on possible
forms of the correlated component.

The ACT results on $\sigma_8$ are also consistent with several recent
studies based on the analysis of ROSAT X-ray flux-selected
clusters. We note that these X-ray cluster studies themselves are
consistent with measures of $\sigma_8$ from richness or weak-lensing
selected cluster samples (see references in the three articles cited
here).  \citet{henry/etal:2009} find $\sigma_8(\Omega_m/0.32)^{0.30} =
0.86 \pm 0.04$ (for $\Omega_m < 0.32$) using cluster gas temperatures
measured with the ASCA satellite.  \citet{vikhlinin/etal:2009} obtain
$\sigma_8(\Omega_m/0.25)^{0.47} = 0.813 \pm 0.013\,({\rm stat}) \pm
0.024\,({\rm sys})$ using temperatures derived from Chandra
observations.  Similarly, \citet{mantz/etal:prep} find that in
spatially flat models with a constant dark energy equation of state,
ROSAT X-ray flux-selected clusters yield $\Omega_m=0.23\pm0.04$,
$\sigma_8=0.82\pm0.05$.

\vspace{1em}
After this article was completed, WMAP 7-year measurements of the
brightness of Mars and Uranus were released
\citep{weiland/etal:inprep}.  They suggest that $T_\mathrm{U}\approx
107\,\kelvin$, which is $\sim 4\%$ dimmer than the value used here.
If the new Uranus temperature is adopted, then all absolute
temperatures and source brightnesses in this work would be reduced by
4\%, and all $C_\ell$ values by 8\%.

\section{CONCLUSIONS}

We have presented the first map and the power spectrum of the CMB sky
made using data from the Atacama Cosmology Telescope at \arone. 
With this map we can compare to WMAP at degree angular scales and
measure point sources with a resolution of 0\fdeg02.  With an unbiased
estimator, we extract the power spectrum, $C_\ell$, over a range of
power exceeding $10^4$.

We have interpreted the spectrum with a simple model composed of the
primary CMB, a possible SZ contribution, and uncorrelated point
sources.  This analysis provides a new upper bound on the SZ signal
from clusters ($\sigma_8^\mathrm{SZ}<\maxSigmaEight$ at 95\% CL, though this is
subject to uncertainty in the SZ models). A coordinated
program of X-ray, optical, infrared, and millimeter-wavelength
observations of the largest SZ clusters is underway.

These high angular resolution measurements probe the microwave power
spectrum out to arcminute scales.  Above $\ell\sim2500$, the spectrum
is sensitive to non-linear processes such as the formation of galaxy
clusters and dusty galaxies.  On the low-$\ell$ end, the spectrum
measures the Silk damping tail of the CMB which can be computed using
linear perturbation theory as applied to the primordial plasma.  It is
clear that to understand the $\ell\gtrsim 1000$ end of the primary CMB, and
thus to improve significantly on measurements of the scalar spectral
index and its running, source modeling will be required. Future
analyses of the ACT data will include the two higher-frequency
channels and additional sky coverage.  In another approach, one can
measure the high-$\ell$ E-modes, because the polarized CMB to
foreground ratio is expected to be higher than that for the
temperature.  We are pursuing both programs, as a
polarization-sensitive camera is currently under development.

\acknowledgements

The ACT project was proposed in 2000 and funded January 1, 2004. Many
have contributed to the project since its inception. We especially
wish to thank Asad Aboobaker, Christine Allen, Dominic Benford, Paul
Bode, Kristen Burgess, Angelica de Oliveira-Costa, Peter Hargrave,
Norm Jarosik, Amber Miller, Carl Reintsema, Felipe Rojas, Uros Seljak, Martin
Spergel, Johannes Staghun, Carl Stahle, Max Tegmark, Masao Uehara,
Katerina Visnjic, and Ed Wishnow. It is a pleasure to acknowledge Bob
Margolis, ACT's project manager. Reed Plimpton and David Jacobson
worked at the telescope during the 2008 season. ACT is on the
Chajnantor Science preserve, which was made possible by the Chilean
Comisi\'on Nacional de Investigaci\'on Cient\'ifica y Tecnol\'ogica.
We are grateful for the assistance we received at various times from
the ALMA, APEX, ASTE, CBI/QUIET, and NANTEN2 groups.  The ATCA team
kindly provided the positions of their 20\,GHz sources prior to
publication.  The PWV data come from the public APEX weather website.
Field operations were based at the Don Esteban facility run by
Astro-Norte.  This research has made use of the NASA/IPAC
Extragalactic Database (NED) which is operated by the Jet Propulsion
Laboratory, California Institute of Technology, under contract with
the National Aeronautics and Space Administration.  We thank the
members of our external advisory board---Tom Herbig (chair), Charles
Alcock, Walter Gear, Cliff Jackson, Amy Newbury, and Paul
Steinhardt---who helped guide the project to fruition.

This work was supported by the U.S. National Science Foundation
through awards AST-0408698 for the ACT project, and PHY-0355328,
AST-0707731 and PIRE-0507768. Funding was also provided by Princeton
University and the University of Pennsylvania.  The PIRE program made
possible exchanges between Chile, South Africa, Spain and the US that
enabled this research program.  Computations were performed on the GPC
supercomputer at the SciNet HPC Consortium.  SciNet is funded by: the
Canada Foundation for Innovation under the auspices of Compute Canada;
the Government of Ontario; Ontario Research Fund -- Research
Excellence; and the University of Toronto.

VA, SD, AH, and TM were supported through NASA grant NNX08AH30G.  ADH
received additional support from a Natural Science and Engineering
Research Council of Canada (NSERC) PGS-D scholarship. AK and BP were
partially supported through NSF AST-0546035 and AST-0606975,
respectively, for work on ACT\@.  HQ and LI acknowledge partial support
from FONDAP Centro de Astrof\'isica.  RD was supported by CONICYT,
MECESUP, and Fundaci\'on Andes.  ES acknowledges support by NSF
Physics Frontier Center grant PHY-0114422 to the Kavli Institute of
Cosmological Physics. KM, M Hilton, and RW received financial support
from the South African National Research Foundation (NRF), the Meraka
Institute via funding for the South African Centre for High
Performance Computing (CHPC), and the South African Square Kilometer
Array (SKA) Project.  JD received support from an RCUK Fellowship.  RH
received funding from the Rhodes Trust.  SD 
acknowledges support from the Berkeley Center for Cosmological
Physics.  YTL acknowledges support from the World Premier
International Research Center Initiative, MEXT, Japan.  The data will
be made public through LAMBDA (http://lambda.gsfc.nasa.gov/) and the
ACT website (http://www.physics.princeton.edu/act/).

\ifnum\manuscript=1
\pagebreak
\input{spectrum_clean.bbl}
\else
\bibliography{../bibtex/act}
\fi

\end{document}